\documentclass[11pt]{article}
\pdfoutput=1
 \usepackage{mciteplus}
 \usepackage{tikz}
 \usepackage{color}
 \definecolor{darkblue}{rgb}{0.1,0.1,.7}
 \usepackage[colorlinks, linkcolor=darkblue, citecolor=darkblue, urlcolor=darkblue, linktocpage,hyperfootnotes=false]{hyperref} 
\usepackage{epsfig}
\usepackage{graphicx}
\usepackage{cite}
\usepackage{amsfonts}
\usepackage{amssymb}
\usepackage{bm}
\usepackage{latexsym}
\usepackage{amsmath}
\numberwithin{equation}{section}
\def\bq{\begin{quote}}
\def\eq{\end{quote}}
 at 10truept

\newcommand{\calo}{{\cal O}}
\newcommand{\calh}{{\cal H}}
\newcommand{\beq}{\begin{equation}}
\newcommand{\eeq}{\end{equation}}
\newcommand{\beqa}{\begin{eqnarray}}
\newcommand{\eeqa}{\end{eqnarray}}
\newcommand{\bea}{\begin{eqnarray}}
\newcommand{\eea}{\end{eqnarray}}

\def\lesssim{~\mbox{\raisebox{-.6ex}{$\stackrel{<}{\sim}$}}~}
\def\roughly#1{\raise.3ex\hbox{$#1$\kern-.75em\lower1ex\hbox{$\sim$}}}

\begin{document}

\thispagestyle{empty}
\begin{titlepage}
  \bigskip

  \bigskip\bigskip

  \bigskip

\begin{center}
{\Large \bf {The deepest problem: some perspectives on quantum gravity}}
    \bigskip
\bigskip
\end{center}

  \begin{center}

 \rm {Steven B. Giddings\footnote{\texttt{giddings@ucsb.edu}} }
  \bigskip \rm
\bigskip

{Department of Physics, University of California, Santa Barbara, CA 93106, USA}  \\
\rm

  \bigskip \rm
\bigskip
 
\rm

\bigskip
\bigskip

  \end{center}

\vspace{3cm}
  \begin{abstract}

Quantum gravity is likely the deepest problem facing current physics. While traditionally associated with short distance nonrenormalizability, it is evident that the long distance problem of unitarity, arising at high energies with black hole formation, is more profound. This reveals a conflict between foundational principles of quantum field theory: those of quantum mechanics, relativity, and locality. Difficulties modifying quantum mechanics suggest a ``quantum-first" approach, with other principles as mathematical properties of a quantum space of states. A challenge is how to describe locality, in terms of Hilbert space structure. Perturbative gravity gives clues, with structure apparently different than in field theory. The mathematical structure of subsystems plausibly supplants conventional locality and plays a foundational role in the theory. This view apparently differs from one of spacetime ``emerging" from entanglement and/or other properties of another quantum system; instead such structure would be more basic. If a black hole behaves as a subsystem, a ``black hole theorem" says that unitarity requires interactions with its environment depending on its quantum state, or more drastic phenomena. Minimal interactions can be parameterized, in an effective approach; they could arise from fundamental dynamics or perhaps wormholes. These or other near-horizon modifications potentially alter electromagnetic or gravitational signatures of this strong gravity region, now being probed in a new era of observation; it is important to seek observational clues for or constraints on such scenarios. A complementary way to investigate quantum gravity is via the gravitational S-matrix.  New perturbative structure has been discovered there, but the harder question again goes beyond to the nonperturbative regime of black hole formation. Associated long-distance behavior of amplitudes indicate novel kinds of analytic behavior, whose further exploration may also provide important clues.  Other key questions regard quantum description of cosmologies, and of associated observables and their relation to observation. The challenge of quantum gravity suggests that maintaining a continued focus on such conceptual and physical, as well as mathematical, questions, and exploring possible observational tests, should remain a high priority. 

 \medskip
  \noindent
  \end{abstract}
\bigskip \bigskip \bigskip 

  \end{titlepage}

\section{Introduction: the problem}

The problem of quantum gravity is probably the deepest problem facing current physics.  In short, we observe that to an excellent approximation, the world is governed by quantum mechanics, and  that long-distance properties of gravity are governed by general relativity, yet there is no known reconciliation of these two frameworks.  Moreover, most observed physics is described to high precision by the Standard Model, and notable exceptions (dark matter, neutrino masses) are readily incorporated in simple quantum field theory extensions of it.  Our goal is to find a reconciliation of the properties of gravity with quantum principles, capable of accommodating -- or better, even explaining --  basic features of the quantum field theory of the Standard Model and possible extensions.  These should arise as a good 
approximation to a more basic theory, and this would provide a foundation for the rest of physics.

While it is naturally tempting to postulate that gravity is just another type of quantum field theory, that view has encountered serious obstacles.  And since quantum field theory can be viewed as a {\it solution} to the problem of reconciling a basic set of principles, these problems strongly indicate that we need to revisit and revise  basic principles in order to give a fundamental description of quantum gravity.  We find multiple clues for this revision in known and expected features of gravity.

The principles underlying local quantum field theory (LQFT) are those of quantum mechanics (QM), special relativity, and locality.  A ``folk theorem" that is sometimes underappreciated is that LQFT is the result of reconciling these principles. This ``theorem" depends somewhat on their precise formulation, particularly that of locality.  For example, Weinberg's text\cite{WeinQFT} argues that LQFT follows from QM, relativity, and cluster decomposition.  Likewise in his QFT lectures\cite{ColeQFT}, Coleman argued that quantum fields result from implementing locality as Einstein causality, in the form of observables commuting outside the light cone.
This ``theorem" appears to express the inevitability of LQFT at low energies, at least in the guise of effective field theory, given the apparent validity of the underlying principles in that regime -- and this explains the remarkable observed success of field theory.

A possible view is that these principles -- and the primacy of field theory -- extend to arbitrarily short distances, which can be explored via the high energy limit.  That motivates this limit as a key way to explore the fundamental properties of the theory of nature.  A first obstacle to this proposal was the discovery of the nonrenormalizability of Einstein gravity, but an obvious counterproposal was that perhaps the structure or principles behind the theory just needed to be modified  somewhat at very short distances, say of order the Planck length $\ell_{Pl}\sim 10^{-33} cm$.  This view informed development of many of the currently dominant approaches to quantum gravity -- loop quantum gravity, string theory, {\it etc.}

But it has become increasingly clear that in gravity there is another, more profound, problem and that local/short distance modifications don't seem to address this problem.  Connected to this is the realization that high-energy scattering -- while still likely probing fundamental properties of the theory -- appears to cease probing short distances, and ultimately involves increasingly {\it long} distance physics.  Moreover, attempts to formulate precise locality in gravity encounter significant subtleties, through closely related behavior.  The ramifications of these appear central to the problem of quantum gravity, and formulation of a new set of principles that govern it.

The apparently more profound problem is that of high-energy unitarity.  This can be motivated by observing that relativistic invariance, together with a very weak version of locality, tell us that we can give relative boosts to two particles which are asymptotically far apart, such that they have an arbitrarily large center-of-mass energy $E$, and we can consider situations where these particles later collide and ask what the theory predicts.  In the classical approximation, for impact parameters $b\lesssim R(E)$, with $R(E)$ being the Schwarzschild radius associated with $E$, the collision forms a black hole\cite{EaGi}.\footnote{For previous arguments in this direction, see {\it e.g.} \cite{Penr,BaFi}.}
The large Schwarzschild radius indicates small curvature at its horizon, and this is traditionally argued to imply that quantum corrections are small.  However, quantum effects are ultimately expected to lead to the black hole's decay, with a leading description of the decay given by the Hawking process\cite{Hawk}.  But, this process continually builds entanglement between the outgoing radiation and the black hole state, and so if the black hole disappears at the end of evaporation,\footnote{The alternative of leaving a microscopic remnant apparently leads to more serious trouble; see \cite{Pres,WABHIP,Susstrouble}.} there is a gross violation of unitarity.  
For a black hole of initial area $A(E)$, this violation is characterized by the missing information 
$\Delta I\sim S_{BH}(E)$, with 
\beq
S_{BH}(E) = \frac{A(E)}{4G}
\eeq
the Bekenstein-Hawking entropy.  There are no obvious quantum corrections that remedy this situation, and moreover such corrections appear forbidden by locality. This gives what is commonly called the black hole information problem, 
 but perhaps a more apt name, recognizing its genericity and the predicament it creates for physics, is the {\it unitarity crisis}.  It appears to play a central role in the problem of quantum gravity.

While assumptions about locality help foster this predicament, further thought reveals that locality itself is remarkably subtle in gravity.  For example, the statement of cluster decomposition must be modified in gravity\cite{SGErice}.  If we consider two arbitrary clusters of excitations scattering in different regions of space, the quantum amplitudes {\it do not} factorize into amplitudes associated to the two separate clusters -- for excitations with sufficiently high total center of mass energy, the clusters are inside a mutual black hole, which ultimately has a dramatic effect on their scattering.  Or, one sees modifications to locality when considering quantum operators generalizing LQFT observables.  These must be gauge invariant, and one way to perturbatively achieve this is to ``gravitationally dress" operators of an underlying field theory --  but then the operators generically do not commute at spacelike separation\cite{SGalg,DoGi1,DoGi2}.\footnote{Earlier work on these effects includes \cite{Heem} and  \cite{KaLigrav}.   The first derived nontrivial commutators as arising from the constraints, but didn't give the dressed operators, and the second focussed on deriving {\it commuting}
bulk operators.}  In this sense even the long-distance, perturbative quantum behavior of the gravitational field seems to signal a precursor to  fundamental modification of locality.

In short, gravity appears to require modifications to traditional locality -- and local or short-distance modifications to physics don't seem to address the most profound difficulties of gravity.  Incorporating these observations into  physics appears to call for a new formulation of basic principles, and it is likely important to maintain a broad and physical perspective in seeking these principles.

In this context, how do the currently dominant paradigms -- which were strongly motivated by short-distance issues -- fare?  

Loop quantum gravity involves significant modification of spacetime structure at near-Planck distances, in particular breaking Poincar\'e symmetry.  This leaves the question of how to recover our approximately Poincar\'e invariant Minkowski vacuum, and low-energy excitations about it, in this approach.  If one can, it is not clear how  long-distance modifications that are sufficient to resolve the unitarity crisis arise; some LQG practitioners have suggested that the resolution is microscopic remnants\cite{HoSm,Pere,BCDHR}, but it is not clear how to avoid the disasterous instabilities of \cite{Pres,WABHIP,Susstrouble}.

String theory also significantly alters the short-distance structure of physics, due to the small but nonzero string size.  This has lead to great success in addressing nonrenormalizability, by removing infinities.  In addition, it has lead to a beautiful, intricate, and mathematically rich structure, that also describes suggestive extensions of the Standard Model.  But, important issues remain.  It is not clear how the extended property of strings addresses the long-distance unitarity problem we have described; strings have behavior that is in a sense too local\cite{LQGST,GGM}.    Moreover, string theory has faced significant challenges in finding phenomenologically  viable extensions of the Standard Model, and in making precise predictions about their phenomenology, resorting instead to a huge and difficult-to-understand landscape of vacua.  The closest string-related ideas have come to addressing the unitarity question is through the proposed AdS/CFT correspondence\cite{Malda}.  But this also has left significant questions unanswered -- some of which will be touched on below -- and specifically the question of how to provide a precise map between boundary theory and the bulk Hilbert space and observables.  Whether or not string theory  ultimately resolves these difficult problems, it is important to investigate other general features of a theory of quantum gravity.

Quantum gravity is {\it one problem}, appearing to call for new principles -- and may well have one solution.\footnote{For example, there is apparently only one correct theory of the strong interactions -- QCD, and likewise for the electroweak interactions.}  We should certainly bear this in mind, as we investigate different approaches to this problem.  But,  given the difficulty and subtleties of the problem, it  does appear very important at the present to maintain a broad perspective on its possible solution, and to pursue a diversity of approaches, rather than placing all of our focus on one or the other particular mathematically-attractive formalism.  It particularly seems important to maintain a continued focus on conceptual and physical questions and tests, as well as possible observational tests, of quantum gravity, and on their relationship to an underlying mathematical description.

\section{A {\it quantum-first} approach, and localization of information}

Since LQFT apparently requires modification, and follows from a set of principles, let's begin by returning to those principles and considering their status.  

Quantum mechanics is increasingly well tested.  Moreover, its mathematical structure is remarkably rigid.  One way to see this is to consider possible modifications.  These appear to typically lead to disaster.  For example, generalization to allow nonunitary evolution, such as more general linear evolution of the density matrix\cite{Hawk-incoh}, generically leads to catastrophic violation of energy conservation\cite{BPS}.  Another alternative is nonlinear generalization of quantum mechanics (see, {\it e.g.} \cite{WeinNL}), but this is found to lead to apparently disastrous communication between branches of the wavefunction and nonlocalities\cite{Polphone}.  Quantum mechanics appears difficult to sensibly modify; if it {\it were} to be modified, an important question is what could replace it.

Relativistic invariance has been tested and found to be accurate to very high precision.  Specifically, while  gravitational fields produce long range curvature, space is approximately locally Min\-kows\-kian, and the Poincar\'e invariance of physics has been tested to great accuracy.  This  includes  for propagation over very long distances, in weak gravitational fields, where modifications to Poincar\'e invariance have been increasingly precisely probed, {\it e.g.} by testing possible modifications to dispersion relations.\footnote{For a recent test and further references, see \cite{Caoetal}.}  Moreover, as general relativity is increasingly tested in strong gravitational contexts, we gain additional constraints on possible violations of relativistic invariance.

In the absence of gravitational fields, locality is also strongly constrained.  This is partly due to the great success of LQFT, and also due to the fact that simple violations of locality generically lead to acausality, and resulting inconsistency.  However, as was pointed out above, we face a challenge of even sharply formulating locality in a theory of gravity, and moreover basic properties of gravity appear to violate formulations of it that have been fundamental to LQFT.  Thus locality appears remarkably subtle in a theory of quantum gravity.

Given the apparent rigidity of quantum mechanics, a reasonable view is that a theory of nature must be governed by quantum principles, sufficiently generally formulated.  This last point is important, since some common assumptions, such as a hamiltonian generating time evolution, require careful examination, and even the notion of spacetime histories\cite{Harthis} is in question.  However, basic principles appear to be those of describing physics by a complex linear space of states with an inner product -- essentially, a Hilbert space --  and the existence of linear operators on this space corresponding to observables, together with unitarity in appropriate contexts.  The role of these as basic principles was considered in \cite{UQM}, which argued they were central to a ``universal" description of quantum mechanics.

This viewpoint of quantum primacy has important implications.  Of course, one could continue to seek modifications of quantum mechanics, and it remains possible that these are needed for gravity.  But, by putting quantum mechanics first, one is accepting a large amount of rigidity of structure.  Specifically, if one adopts this viewpoint, the other principles need to be formulated as properties of a theory of Hilbert space states.  Applied to quantum gravity, this is the foundation of what has been called the ``quantum-first" approach\cite{QFG,QGQF}.  Here, the relative rigidity of quantum mechanics is an important constraint, which can be combined with multiple clues about the structure of gravity, and of course the necessity of embedding Standard Model physics.

The theory of quantum gravity appears to have states that are very close to Minkowski space, at scales short as compared to ambient curvature radii.  A reasonable (though not strictly necessary) postulate is that it has a state $|0\rangle_M$ that is the corresponding quantum state, as well as perturbations about it -- although our cosmos is in a different state.  If so, special relativistic invariance is straightforward to describe.  Specifically, there should be an action of the Poincar\'e group that leaves $|0\rangle_M$ invariant, and acts in the familiar way such that excited states correspond to representations of it, and quantum amplitudes such as those of the S-matrix transform appropriately.  Similar statements might be extended to the contexts of anti de Sitter space or de Sitter space.

The challenge is that of locality.  In view of our quantum-first approach, this should be formulated in terms of structure on the Hilbert space.  Localization is easy to describe for example in familiar finite-dimensional Hilbert spaces.  If a Hilbert space has a factorization, 
\beq\label{Hfac}
\calh =\calh_1\otimes \calh_2\ ,
\eeq
then information, observables, {\it etc.} may be thought of as localized in one or the other factor, in an obvious way.  However, in LQFT, this already becomes more subtle.  The basic quantum operators are fields, which can be localized in different regions of space, for example by integrating against compact-support test functions.  However, this does {\it not} imply a factorization of the Hilbert space \eqref{Hfac} corresponding to neighboring regions of space, due to the type III property of operator algebras of field theory.  This can colloquially be thought of as due to infinite entanglement of short distance modes, whose existence is for example  implied by the Poincar\'e invariance of the theory.

The LQFT remedy for this is to instead consider properties of the algebra of observables; this is basic to the algebraic approach to QFT.\footnote{See, {\it e.g.}, \cite{Haag}.}  Subalgebras may be associated to composite operators of fields corresponding to one or another spacetime region, and a fundamental postulate, implementing locality, is the postulate that these subalgebras commute for spacelike separated regions.  The structure defined on the Hilbert space by such a ``net" of subalgebras captures the topological and causal structure of the underlying spacetime manifold -- it encodes this manifold structure in the quantum theory.  In this sense, locality is a property that is hardwired into the Hilbert space at the outset.

Gravity adds significant subtlety.  One can start with a local operator of an underlying LQFT, say that of a scalar field $\phi(x)$.  However, this is not gauge invariant under the diffeomorphism symmetry of general relativity, which can be thought of as generated by the constraints
\beq\label{constraints}
C_\mu(x)=  \frac{1}{8\pi G} G_{0\mu}(x)-T_{0\mu}(x) =0\ ,
\eeq
with $G_{\mu\nu}$ the Einstein tensor and $T_{\mu\nu}$ the energy-momentum tensor (including possible cosmological term).  
A perturbative approach to making a gauge-invariant operator ``as local as possible" in the low-energy theory is to {\it gravitationally dress} an underlying operator, {\it e.g.} defining
\beq\label{dressop}
\Phi_V(x^\mu) = \phi(x^\mu + V^\mu(x))\ ,
\eeq
or, for a more general underlying operator $\calo$, to leading order in $\kappa = \sqrt{32\pi G}$,
\beq\label{gendressop}
\hat \calo \simeq e^{i\int d^3x V^\mu(x)T_{0\mu}(x)}\calo e^{-i\int d^3x V^\mu(x)T_{0\mu}(x)}\ .
\eeq
Here $V^\mu(x)$ is a functional of the metric perturbation, defined by expanding about a background $g_{\mu\nu}$, via $\tilde g_{\mu\nu}= g_{\mu\nu}+\kappa h_{\mu\nu}$; such leading-order dressings can be found\cite{DoGi1,GiKi,QGQF,DoGi4,HaOo,GiWe} so that $\Phi_V$ or $\hat \calo$ commute with the constraints \eqref{constraints}.  An example is the line dressing\cite{DoGi1,QGQF}
\beq
V_\mu^L(x)= \frac{\kappa}{2} \int^\infty_x dx^{\prime\nu} \left\{ h_{\mu\nu}(x') +
 \int^\infty_{x'} dx^{\prime\prime\lambda} \left[ \partial_\mu h_{\nu\lambda}(x'') - \partial_\nu h_{\mu\lambda}(x'')\right]\right\}\ .
\eeq
These dressings are non unique; 
another is a spherically-symmetric Coulomb dressing, yielding linearized Schwarzschild\cite{DoGi1}.  In general the difference between such possible dressings corresponds to addition of
radiative (sourceless) gravitational fields.  

Although so far these are only low-energy and perturbative constructions, they suggest important implications for the more complete nonperturbative structure.   
 
A first consequence is that the dressed observables based at different points $x$ and $y$ don't generically commute for spacelike $x-y$.  The dressing, which has the interpretation of creating the gravitational fields of the underlying field quanta, leads to nontrivial commutators.  These are typically proportional to the gravitational potentials of the quanta\cite{DoGi1}.  

The algebras of quantum observables in gravity as a result have a much different and more intricate structure than those of LQFT\cite{SGalg,DoGi1}.  It is then not clear how to use them to define locality.  There is also a question of how to connect these quantum observables to actual observation.

Locality played a key role as a foundational principle of LQFT.  In fact, in that context, generic nonlocality appears nonsensical, in that it leads to inconsistency when considering perturbations about a Minkowski background.  The  argument is that noncommutativity of observables at spacelike separations allows signaling or transfer of information at these separations.  A Lorentz boost can convert this into signaling backward in time, and two such signaling processes can be combined to allow an observer to signal into their past lightcone.  This leads to inconsistencies such as the ``grandmother paradox," where an observer sends a signal designed to cause the death of their grandmother before their mother is born.

More generally, a notion of localization of information plays a key role in interpreting physics, say by separating observer and observed\cite{EinsSep,Howa-Eins}.  And, this localization underlies various 
useful quantum information-theoretic characterizations of physics, such as definitions of entropies, discussion of transfer of information, {\it etc.}. 

So, a key question is what replaces localization of information, and other aspects of locality, in quantum gravity?  The comparison with LQFT suggests that this may be crucial for interpretation and consistency of the theory.

Specifically, a starting point appears to be  the question of how to define {\it subsystems} in quantum gravity.  From the quantum-first perspective, these must be defined in terms of a mathematical structure on the Hilbert space.  However, unlike with LQFT, this structure is not obviously furnished by the algebra of quantum observables.  A definition of subsystems appears to be a prerequisite for precise information-theoretic discussions, and plausibly -- as with LQFT -- plays a key role in the foundational mathematical structure of the theory.

The question of defining subsystems -- or a suitable approximation -- is a general one for the nonperturbative theory.  But there are hints and  possible precursors  in the perturbative structure of the Hilbert space.

We begin with the question of localization of information.  First, in an underlying LQFT that we are to couple to gravity, consider an operator $\calo_x$ localized with compact support in a neighborhood of point $x$, {\it e.g.} $\calo_x = \int d^4x' J_x(x') \phi(x')$, with compact support function $J_x$, and consider the state
\beq
|\psi\rangle_x=e^{i\calo_x}|0\rangle\ .
\eeq
In LQFT, there is no way that measurements at point $y$ spacelike separated from the neighborhood can differentiate this state from the vacuum:
\beq
{}_x\langle\psi|\calo_y|\psi\rangle_x =  \langle 0|e^{-i\calo_x} \calo_y  e^{i\calo_x}|0\rangle = \langle 0|\calo_y|0\rangle 
\eeq
for general such spacelike separated observables $\calo_y$.  

Coupling to gravity will modify this.  Specifically, with some dressing $V^\mu(x)$, consider the dressed state (to leading order), corresponding to \eqref{gendressop}, 
\beq\label{dstate}
|\widehat \psi\rangle_x \simeq e^{i \int d^3x' V^\mu(x') T_{0\mu}(x')} |\psi\rangle_x ,
\eeq
 and annihilated by the constraints.\footnote{More carefully, by half of the constraints\cite{DoGi1}, as in Gupta-Bleuler quantization\cite{Gupt}.}
General expectation values of spacelike separated operators that include the metric perturbation $h_{\mu\nu}$ {\it can} distinguish this state from the vacuum, since the dressing creates a non-zero gravitational field that must extend to infinity\cite{DoGi2}.  

However, here the non-uniqueness of the dressing, corresponding to shifts by radiative fields, plays an important role.  A dressing may be chosen\cite{DoGi4} so that for different such states \eqref{dstate}, the dependence of the asymptotic gravitational field is only through the Poincar\'e charges, or more general moments of these charges, for the the state\cite{SGsplit}; for further discussion see \cite{SGasymp}.  Essentially, this is a quantum generalization of classical results that the gravitational field far outside a matter distribution may be, for example, taken to be simply the boosted Kerr solution\cite{CoSc,ChDe}.\

This means that if there are different states in a neighborhood with identical Poincar\'e moments, those are not distinguished by distant measurements of their perturbative gravitational dressing.  In this sense there can be localization of information; this construction was referred to as a perturbative gravitational splitting in \cite{SGsplit}.\footnote{This result also argues against\cite{DoGi3,DoGi4,SGsplit} a role for ``soft charges"\cite{Hawk-info,HPS1,HPS2} in resolving the information problem.}

However, at the nonperturbative level, there are arguments that distant observables can detect the differences between dressed states\cite{MaroUH,JacoBU}\cite{DoGi3}.  
These arise because in gravity the momentum operators can be written as
\beq
P_\mu = P_\mu^{ADM}[h(\infty)] + \int d^3x\  C_\mu(x)
\eeq
with $P_\mu^{ADM}$ the asymptotic ADM expressions, and thus become purely asymptotic operators acting on operators or states annihilated by the constraints.  As in \cite{DoGi3} one can consider expressions where the momentum operators translate the state into the asymptotic region, where another asymptotic observable can detect it:
the expectation value of asymptotic observables
\beq
{}_x\langle\widehat\psi|\calo_y e^{iP_i^{ADM}c^i}|\widehat \psi \rangle_x
\eeq
depends on details of the underlying state, for certain observables $\calo_y$ and translations $c^i$.  This requires  a large translation, and consequently construction of the dressing beyond leading order.

This complicates the general question of how to describe localization of information and define subsystems.  One possible view is that there is no basic localization of information in quantum gravity.  In fact a related but more complicated argument was suggested to explain AdS holography in \cite{MaroUH,JacoBU}, by mapping a ``localized" state $|\widehat \psi\rangle_x$ at time $t$ to the AdS boundary via evolution, then back to $t$ by translation via $P_0^{ADM}$.  However, there are also apparently significant subtleties in such arguments for holography\cite{JaNg,SGHolo}.   These involve both the closure of the boundary algebra, and particularly the need to solve the nonperturbative analog of the constraints as a starting point -- which is tantamount to solving the problem of quantum gravity.

Indeed, it has even been argued\cite{CGPR,LPRS} that a perturbative version of this discussion gives asymptotic access to information.  These arguments were examined in \cite{SGasymp} and found, however, to lead to effects exponentially small in the distance from the neighborhood, which make measurements correspondingly difficult.  

Of course if a complete holographic map can be found to explain AdS/CFT, as is widely believed, that would support a view in which information is fundamentally delocalized in the bulk.  If the correspondence to ${\cal N}=4$ super Yang Mills is precise, there is an underlying localization in the {\it boundary}  spacetime.  However, part of the difficulty is understanding how this localization can be related to the bulk localization in AdS, which holds approximately for weak gravitational perturbations, and exactly in the LQFT limit of vanishing Newton's constant.\footnote{There is a proposal relating bulk subregions extending to infinity to boundary subregions, via a ``subregion-subregion" duality; for a recent overview, see \cite{Boussoetal}. This has been formulated at the semiclassical level, and it is also not clear how it would extend to subregions isolated from the boundary.}  If the appropriate bulk mathematical structure on the Hilbert space {\it can} be realized in AdS/CFT, that would then furnish an example of a quantum-first description of gravity.

In short, a key question remains of how to define the mathematical structure of quantum gravity that is relevant to localization of information.  
We would like to infer  the necessary structure, on a Hilbert space that can be interpreted in terms of bulk quantum gravity states.
We have seen precursor obstacles and hints already at the perturbative level.  Possibly a more intricate construction will lead to a mathematical definition of subsystems that reduces to familiar definitions in the weak gravity limit.  

Another possibility to consider is that such localization is only approximate, say in the limit of weak perturbations of the gravitational field.  In this sense, one might have decompositions of the Hilbert space \eqref{Hfac}, or generalizations\cite{SGsplit} to account for asymptotic measurability of Poincar\'e charges, that are only approximate.
This would require a careful interpretation of such factorization that is only approximate, in describing physics.
And if it is true that such locality is only approximate, then that returns us to the question of what deeper principle replaces it and plays an analogous role in the fundamental structure and consistency of the theory.

It is interesting and apparently important that any such notions of locality apparently suffer significant modification in strong gravitational fields resulting from the high-energy limit, as outlined above.  It likewise seems important that these modifications, which are also associated with the ideas of ``holography," apparently arise from large concentration of energy, rather than large concentrations of information.

Beyond these questions of fundamental structure there are of course further questions in giving such a quantum-first description of physics.  What principles besides unitarity govern evolution on the Hilbert space?  How is such evolution fundamentally defined in states which don't correspond to asymptotically flat or AdS geometries?\footnote{This may be described using a different kind of relational observable, as in \cite{GMH}.}  How are observables defined, and related to a theory of observation?
And, of course, how is structure incorporated that reduces to that of Standard Model fields and interactions, in the weak-gravity limit?

\section{Quantum black holes}

As described above, black holes, or their quantum versions, are expected to govern the generic high-energy behavior of a Poincar\'e-invariant theory of physics, where gravity becomes strong.  They are also apparently ubiquitous in our quantum universe, due to the universality of gravity and the resulting genericity of gravitational collapse.  A theory of quantum gravity needs to explain their evolution, and here, as outlined above, it confronts an apparent crisis of unitarity.  

 This crisis appears directly related to the questions of locality.  Consider a collision of two very high-energy particles in a pure state, at impact parameters where BH formation dominates.  The resulting large BH is expected to have an approximate description via classical geometry. The LQFT approximation to evolution on BH geometries was developed by Hawking\cite{Hawk} and in many subsequent works.  Quanta are emitted that are entangled with corresponding internal excitations that are part of the internal state of the BH.  By locality, the entanglement of the internal excitations cannot be transferred to external excitations, as this would require information transfer outside the light cone.  If the BH disappears at the end of evaporation, that results in a missing information characterized by a von Neuman entropy of size $S_{BH}(E)$, representing an enormous violation of the quantum-mechanical unitarity principle.
 
 In fact, this problem can be stated in an even more general way, in the form of a ``black hole theorem"\cite{BHthm}; this goes beyond the LQFT description, which furnishes an example. This states that if I) a black hole is a subsystem; II) black hole states have identical exterior evolution and III) a black hole disappears at the end of its evolution, then evolution of the BH plus surroundings violates the unitarity principle.\footnote{Further clarification of the assumptions appears in \cite{BHthm}.}  
 
 This statement connects to locality in two ways; first, one needs to describe localization of information in the BH subsystem, and second  in the identical exterior evolution postulate, which for example is implied by locality in the LQFT approximation.  Resolving this crisis while preserving quantum mechanics as well as correspondence with the low-energy validity of LQFT seems to require modification of locality, and so the crisis appears to also give an important guide to the fundamental principles of quantum gravity.
  
Preserving quantum mechanics apparently calls for violation of postulate I or II.\footnote{Violation of postulate III leads to scenarios with microscopic remnants with unbounded degeneracies, and consequent infinite production\cite{Pres,WABHIP} and other\cite{Susstrouble} unacceptable features.}  These violations appear to need to operate on scales of order the BH  horizon size $R$, or larger; {\it microscopic} violations of locality don't appear to lead to the necessary modifications to these postulates.  In fact, a near consensus on this point in much of the quantum gravity community is evident in the many proposals that have been made modifying the description of a BH via classical geometry plus LQFT at such scales.  These include massive remnants\cite{BHMR}, including in the form of gravastars\cite{MaMo}, fuzzballs\cite{fuzzrev}, firewalls\cite{AMPS}, and Planck stars\cite{RoVi}, as well as the proposals of ER=EPR\cite{MaSu}, 't Hooft\cite{thooft}, and replica wormholes\cite{AEMM,AMMZ,PSSY,AHMST,AHMSTrev}.  
 
Such proposals typically involve considerable extra structure that is difficult to consistently describe, or may lack a systematic description.  
Physics often obeys a principle of parsimony, finding the most simple solution to a given problem, and that suggests looking for minimal departures from the postulates  consistent with physics in domains we know, but  sufficient to resolve the crisis.

As described in the preceding section, in gravity it is difficult to precisely characterize subsystems, and so this could be at the root of a possible resolution.  However, the ability to localize information to an excellent approximation, for example in the gravitational splitting construction, along with the important roles of subsystems, argue that a subsystem decomposition might be defined to an approximation that is good enough that this is not the resolution.  While these questions warrant further exploration, let us assume that postulate I holds, to a good approximation.  In that case, for a resolution we should seek a description of violation of postulate II.  

Violation of postulate II means that evolution of the exterior subsystem to a black hole depends on the BH state.  In the LQFT approximation, this is forbidden by locality, since excitations inside a BH can't influence those outside.  But, if we are going to preserve the equivalence principle as best we can, we can describe evolution of the BH plus surroundings as that due to LQFT, plus corrections that are small but  large enough to resolve the problem.  This specifically implies  interactions between the BH state and the external state, with the latter taken to be otherwise well-described by LQFT. 

We will parameterize the subsystem structure by saying that the states take the form $ |\psi_{BH,i}, \psi_{env}\rangle$, where $i$ describe the BH internal states and $\psi_{env}$ describes the independent state of the 
environment.\footnote{$\psi_{env}$ may depend on the Poincar\'e charges of the BH, but by hypothesis not on other aspects of its state.}
In a hamiltonian description of evolution of the combined system,\footnote{Such a hamiltonian description in LQFT has been given for two-dimensional black holes in \cite{SG2d,SEHS}, and will be discussed in higher dimensions in \cite{GiPe}.} the interactions  imply  a correction to the LQFT hamiltonian,
\beq\label{Hpdh}
H= H_{LQFT}+\Delta H\ .
\eeq
We expect this correction has an important piece altering the evolution of the BH state, as well as an interaction term $\Delta H_I$ between BH and environment.
Further considerations motivate the structure of 
$\Delta H_I$.  First, the simplest interaction that transfers information from BH state to environment is bilinear in operators acting on each of the subsystems.  This takes the general form
\beq\label{Genint}
\Delta H_I = \sum_A \lambda_{ij}^A \calo_A^{env}\ ,
\eeq
where $\lambda^A$ gives a basis for operators on the BH subsystem, and $\calo_A^{env}$ are some operators on the environment; in our approximation the latter are taken to be operators constructed in LQFT.

A general question is what such terms are capable of resolving the crisis.  To do so, they must transfer information (specifically, entanglement\cite{HaPe,GiSh1,Sussxfer}) at a rate sufficient to compensate for the entanglement growth of the Hawking process, which implies at $\calo(1)$ qubit per time R.  General conditions for information transfer rates via such interactions, and their relation to coupling sizes, can be studied, as for example in \cite{GiRo}.  

However, we can motivate further refinement of the general expression \eqref{Genint}.  First, for a minimal solution of the problem which doesn't create causality or other problems, we don't expect such operators to transfer information to distant reaches of the Universe.  And, since there are arguments that Hawking radiation is produced in a ``quantum atmosphere" region of size $\sim R$ near the BH\cite{SGBoltz}\cite{SG2d},\footnote{See \cite{Unru,Full,Bard} for earlier related arguments.} it is very plausible for the interactions to likewise be localized in this region.  On the other hand, one could assume that they are highly tuned, to be localized to within a distance, say $\sim \ell_{Pl}$, of the horizon.  This tuning goes beyond the natural scale $R$, and has an unnatural result: production of high-energy particles, for example as seen by an infalling observer, near the horizon\cite{SGTrieste,Brau}\cite{AMPS}, called a ``firewall" in \cite{AMPS}.  

Second, both the universal nature of gravity, and the consequent beautiful aspects of BH thermodynamics, suggest a universal coupling to all fields.  This is also important in
addressing\cite{NVNL,NVUEFT,NVNLT} Gedanken experiments of black hole mining\cite{UnWamine,LaMa,FrFu,Frol}, where BH decay rates can be increased by threading them with a cosmic string or introducing other mining apparatus, and there needs to be commensurate increase of information transfer.  Universality can be achieved by restricting \eqref{Genint} to couple to the stress tensor (including that of the spin two graviton modes)\cite{NVUEFT,NVNLT,NVU},
\beq\label{Tint}
\Delta H_I =\sum_A \lambda^A \int dV G^{\mu\nu}_A(x) T_{\mu\nu}(x)\ .
\eeq
If the ``form factors" $G^{\mu\nu}_A(x)$ are taken to have support and spatial variation at scales $r\sim R$, that respects the preceding naturalness conditions.  They likewise naturally connect states with energy differences $\sim 1/R$.  We view
\eqref{Tint} as describing an effective hamiltonian, parameterizing our current ignorance of the detailed form of the interactions, which might be inferred from more basic principles, once those are understood.

This effective hamiltonian can be reorganized by defining
\beq
H^{\mu\nu}(x) = \sum_A \lambda^A  G^{\mu\nu}_A(x)\ ,
\eeq
and becomes
\beq\label{Hre}
\Delta H_I = \int dV H^{\mu\nu}(x) T_{\mu\nu}(x)\ .
\eeq
This coupling implies $H^{\mu\nu}(x) $ behaves like a perturbation of the metric, but it is also a quantum operator that depends on the BH state.  

To further constrain these couplings, we return to the condition of sufficient information transfer, at rate $\calo(1/R)$.  One way to achieve this\cite{NVNLT} is if in a typical BH state, 
\beq\label{Hst}
\langle \psi_{BH}, t|H^{\mu\nu}(x) |\psi_{BH},t\rangle =\calo(1)\ .
\eeq
This is clear since an $\calo(1)$ metric perturbation with spatial and temporal variation scales $\sim R$ can clearly transfer information at a rate $\calo(1/R)$.  This represents a relatively large variation of the metric near the BH, and is referred to as a strong or coherent scenario.

However, in pursuing a {\it minimal} resolution of the crisis, one can ask what size couplings are {\it necessary} for this information transfer rate.  A simple estimate of the information transfer rate due to a small perturbation to the hamiltonian is that it is comparable to the rate at which the perturbation induces transitions\cite{NVU}, and this can be confirmed in simple examples\cite{GiRo}.  This gives
\beq\label{FGR}
\frac{dI}{dt} \sim \frac{dP}{dt} = 2\pi \rho(E_f)|\Delta H_I|^2
\eeq
by Fermi's Golden Rule, with typical final density of states $\rho(E_f)$ for the combined subsystems, and typical size $|\Delta H|$ for the relevant matrix elements of $\Delta H$.
The density of states should include that of the BH, which is $\exp\{S_{bh}\}$, with $S_{bh}$ expected to be comparable to the Bekenstein-Hawking value $S_{BH}$.  Then, an $\calo(1/R)$ rate can result from exponentially tiny couplings,
\beq\label{DHw}
|\Delta H_I|\sim e^{-S_{bh}/2}\ ,
\eeq
which also corresponds to metric perturbations in a typical state of size
\beq\label{Hwk}
\langle H_{\mu\nu}(x)\rangle = \calo(e^{-S_{bh}/2})\ .
\eeq
This is referred to as a weak or incoherent scenario\cite{NVU}. 

Either the strong or weak scenarios do violate the equivalence principle in the strong gravitational field of a  BH, but do so in a way that is modest and in particular not violent to infalling observers, unlike firewalls\cite{SGTrieste,Brau}\cite{AMPS}, or for example the expectated behavior for fuzzballs.  In this sense they are close to preserving the equivalence principle, and this is particularly true for the weak/incoherent scenario, which has only very subtle effects on infalling observers.

These interactions have been motivated from very general considerations, and are quite minimal.  If they are part of the consistent description of BHs, an important question is to derive them from more basic principles of quantum gravity.  This is left for the future.   In fact, the apparent necessity of such interactions, and their form, may provide important clues about these principles.  Also, effective interactions of the general form \eqref{Genint} might result from some of the proposals in the literature, although some of these, such as fuzzballs\cite{fuzzrev} and firewalls\cite{AMPS}, introduce hard, rather than soft, structure at the horizon, corresponding to much more significant departures from the equivalence principle.

In particular, there has been considerable recent discussion of the replica wormhole approach to ``calculating" BH entropies\cite{AEMM,AMMZ,PSSY,AHMST,AHMSTrev}.  In the end, this essentially just  supplies a prescription to connect the known rising curve of entanglement entropy from Hawking's calculation\cite{Hawk} to the expected falling curve of Bekenstein-Hawking entropy for an evaporating BH, and is based on a formal extrapolation of euclidean quantum gravity methods\cite{GiTu}.
These methods have not yet, however, given a way to calculate the underlying quantum amplitudes for corresponding processes.  There has been a suggestion\cite{MaMa} that they are associated to processes with ``real" wormholes, that correspond to emission of baby universes, as has been considered previously\cite{LRT,Hawkworm,GiStinst}.  Such processes were argued\cite{Cole,GiStinc} to lead to shifts in the effective lagrangian or hamiltonian of our universe, which could in fact be an explanation of couplings such as \eqref{Genint}.  What is not yet explained is the form of these couplings -- which likewise could be nonlocal on the BH scale $R$, corresponding to emission of a baby universe of this size -- and whether they have the correct structure to account for amplitudes with the necessary properties.

Since many scenarios imply modifications at the BH scale $R$, and since we are now able to observationally investigate these scales  via observations of both electromagnetic and gravitational radiation from black holes, that strongly suggests seeking evidence for or constraints on such scenarios\cite{SGObs,SGLIGO}.  This is an important general problem\cite{SGAstro,GKT}.  
Further constraints are needed to do so with general interactions \eqref{Genint}, but the more specific interactions \eqref{Tint} and \eqref{Hre} suggest possible observational signatures and constraints.  

Specifically, if the strong scenario \eqref{Hst} is realized, that is expected to lead to $\calo(1)$ deformation of light trajectories sufficiently close to the horizon, in a time dependent fashion.  Simple models of such effects\cite{GiPs} have shown they can be quite pronounced in light images of BHs, so much so that the current Event Horizon Telescope images of M87 are apparently providing significant constraints on such scenarios.

The weak scenario \eqref{Hwk} is, on the other hand, not expected to strongly influence light propagation\cite{NVU}.  But, it can apparently influence gravitational wave absorption, and thus ultimately could have an effect on gravitational wave signatures.  To see this, consider transition amplitudes where a gravitational wave is incident on a BH with corrected hamiltonian \eqref{Hpdh}.  Now, replacing \eqref{FGR}, we have transition amplitudes of the form
\beq
\frac{dP}{dt}=2\pi\rho(E_f) \Biggl |\int dV \langle i| H^{\mu\nu}|\psi\rangle\langle \beta| T_{\mu\nu}|\alpha\rangle\Biggr|^2\ ,
\eeq
between graviton states $|\alpha\rangle$ and $|\beta\rangle$.  While the matrix elements of $H^{\mu\nu}$ are exponentially small, \eqref{DHw}, again the density of states compensates, and can yield $\calo(1)$ probabilities.  The spatial and temporal variation scales $\sim R$ of $H^{\mu\nu}$ then imply transitions between states differing by momenta or energies $\sim 1/R$.  Frequencies of this size play a role in the gravitational wave signal of a coalescing binary BH.  In particular, these contributions to amplitudes are expected to modify absorption and reflection cross sections, possibly with effects on the gravitational wave signal of the merger.

It is important to explore the possible observational signatures of these as well as of other viable scenarios further, and test them in the newly-founded regime of strong gravity observations.

\section{The gravitational S-matrix}

The preceding sections review arguments suggesting fields and their lagrangians do not necessarily give a fundamental description of the physics of quantum gravity.  Other modern developments also hint in this direction.  One is that of dualities, where different field theories with different lagrangians are believed to lead to equivalent physics.  Another is study of perturbative scattering, where deeper simplicities have been found that are not evident in a straightforward expansion in Feynman diagrams.  

A complementary way to investigate quantum gravity, in states with asymptotic behavior like that of Minkowski or AdS space, is through properties of its scattering amplitudes.  Infrared behavior complicates this in four dimensions.  Two ways around this are to either carefully treat soft modes, or to work in higher dimensions where infrared behavior is improved.  This discussion will mainly focus on the latter approach, with the implicit assumption that quantum gravity can be defined in higher dimensions, although ultimately more careful treatment of soft modes is desired,\footnote{For work in this direction, see {\it e.g.} \cite{astrosrev}.}  but is not expected to change central aspects of the discussion.  The structure of the S-matrix plausibly supplies powerful clues about a theory; for example, string theory can be viewed as originally  {\it discovered} by Veneziano's guess of an S-matrix with certain structure\cite{Vene}.

At the perturbative level, there have been powerful new results in the study of gravitational amplitudes.  These include those of the generalized unitarity methods and the beautiful double copy structure (for a review see \cite{BCetalrev}) where perturbative-level integrands of amplitudes are double copies of the kinematic structure already present in the amplitudes of Yang-Mills.  Moreover, these and related methods have provided a powerful way to study the possible structure of perturbative infinities, and the renormalizability structure of the theory (for recent developments and further references see \cite{Bern:2018jmv}).

However, as was discussed above, while the short-distance problem of nonrenormalizability has historically been perceived as the central problem of quantum gravity, there has been a growing realization that the problem of unitarity is apparently more profound.  The previous discussion has argued this involves important long distance departures from current understanding, rather than simply altering short ({\it e.g.} Planck) distance behavior.

The introduction and preceding section described the importance of understanding scattering in the high-energy limit, where the unitarity crisis is confronted.  
To investigate this, and understand its connection with perturbative scattering amplitudes,
consider high energy scattering as a function  of impact parameter; this is also closely related to a description in terms of momentum transfer  $q$.  We consider this in a general spacetime dimension $D$, with Newton constant $G_D$.  (For a more extensive discussion, see \cite{SGErice}; earlier work in the context of string theory includes \cite{Amati:1987wq,Amati:1987uf,Amati:1988tn}).
Even at ultraplanckian center-of-mass energy $E$, gravitational scattering is very weak and dominated by tree or Born level exchange for sufficiently large impact parameter $b$, specifically as long as $\chi\sim G_D E^2/b^{D-4}\ll 1$.  However, the Born approximation fails as $\chi$ approaches unit size.  For fixed ultraplanckian energy and smaller impact parameters, the sum of ladder diagrams, corresponding to iterated single-graviton exchange, unitarizes the amplitudes\cite{GiSr}; these diagrams of course sum to give the eikonal amplitudes\cite{Dittrich:1970vv},\cite{Amati:1987wq,Amati:1987uf,Amati:1988tn,MuSo,KaOr,Dittrich:2000be}.  

This regime corresponds to classical scattering, analogous to that of macroscopic objects like planets. It moreover begins to illuminate some important new features.  Specifically, the scattering is dominated by a saddlepoint, with total transverse momentum transfer given by $q_\perp\sim \partial \chi/\partial b$.  While this momentum transfer can also be superplanckian, an important feature of gravity is that of {\it momentum fractionation}\cite{GSA}:  the total momentum transfer is carried by the large number of graviton exchanges\cite{GiPo}, such that each exchanged graviton carries a typical momentum $k\sim 1/b$ that is subplanckian.  

This suggests some important conclusions\cite{GSA}.  First, even for very high ultraplanckian momentum transfers, it is not the properties of individual diagrams that governs the scattering behavior; rather it is the sum of diagrams that yields the dominant saddlepoint. Secondly, the momentum transfer $k\sim 1/b$ of the individual exchanged gravitons indicates that the process is only probing large distances $\sim b$ -- not subplanckian distances, despite the superplanckian $E$ and total $q$.  This is not strongly dependent on the short-distance structure of individual diagrams.  These two observations combined add further support to the view that 
renormalizability -- involving the short distance behavior  of diagrams -- does not play a central role in this ultraplanckian regime.  The high energy behavior  emphasizes the need to understand properties of {\it sums} of diagrams or their nonperturbative extension, not just individual diagrams.

While radiation and other effects can also begin to make important contributions at decreasing impact parameter, particularly important effects occur in the vicinity of the impact parameter $b\sim R(E)\sim (G_D E)^{1/(D-3)}$.  Here, as outlined, classical scattering leads to black hole formation\cite{EaGi}, and one can argue that this classical process provides a starting point for a semiclassical treatment of the scattering.  At the diagrammatic level, this regime is associated with a sum of a new class of diagrams becoming important, those consisting of graviton tree diagrams inserted between the high-energy legs.  In fact, Duff\cite{Duff} has shown in the simple example of Schwarzschild that summing such diagrams builds up the black hole geometry.  

This introduces a new set of problems.  First, the perturbative description of the geometry fails at the horizon; the perturbation series appears to {\it diverge}.  One can nonetheless attempt to treat the scattering in a semiclassical expansion about the classical geometry -- similar to one about the eikonal saddle.  But, quantization of fluctuations about this geometry leads to the failure of unitarity\cite{Hawk} that we have described.  We expect 
 that restoration of unitarity and resolution of the crisis involves new intrinsically quantum-gravitational effects, and an important question is 
 whether clues inferred from studying the S-matrix for gravity can help shed light on these effects. 

Some may expect string theory to make important modifications.  However, in a sense, perturbative string theory appears ``too local" to do so; the extended nature of the string means that strings are tidally excited\cite{LQGST,GGM} in such collisions, but the resulting excitations appear confined within the classical horizon and don't appear to furnish the necessary long-distance modifications to evolution.

A key question, then, is what can be inferred about the structure of amplitudes in this regime.  We would like to understand how the basic mathematical and physical structure of quantum gravity manifests itself in the full amplitudes.
Analyticity, together with other constraints, can be very powerful, as the discovery of string theory exhibited\cite{Vene}.  Further exploration of this question seems important, and potentially fruitful.

Since we expect gravity modifies the locality of LQFT, it is important to probe this question and its role in amplitudes.    Of course the low energy behavior of amplitudes exhibits subtleties and structure familiar from other massless theories.
Questions such as locality and factorization of the Hilbert space are more obscured in the S-matrix approach, but connect with important properties.  While polynomial boundedness of amplitudes in certain regions of complex Mandelstam $s$ is important for causality, there is in particular evidence that the long-range nonlocal properties of quantum gravity leads to 
certain nonpolynomiality of amplitudes\cite{GiPo},\cite{SGErice}. This is already seen at the level of the nonperturbative eikonal amplitudes.  Incorporating anticipated contributions from black holes\cite{GiSr}\cite{GiPo}\cite{SGErice} suggests different nonpolynomial behavior, that may also not be polynomially bounded in analytically continued regions.  Such novel analytic structure, and its connection to locality and its possible modifications, is important to understand.  Possibly deeper understanding, together with study of other aspects of analytic behavior, can furnish important clues, or even suggest inspired guesses for the S-matrix in the spirit of \cite{Vene}.  It is important to further study and characterize these analytic properties of gravitational scattering and their constraints.

Another important question is that of whether new aspects of gravitational amplitudes that are found perturbatively, such as the double copy structure and color-kinematics duality, are a part of a deeper structure that also governs the behavior of perturbative sums of amplitudes and extends to the nonperturbative regime.  
A possible counterpoint is that some of the mathematical structure may be associated with the redundancies and symmetries of calculational approaches; for example in string theory, there are various nice properties of triangulations of moduli spaces\cite{GMW,GiWo}, and corresponding amplitudes, but these don't necessarily connect to intrinsic properties of moduli space, or go beyond to the nonperturbative regime.  Though, hints that this could be the case are found in discoveries of double copy structure for classical solutions, such as that for the Schwarzschild black hole\cite{Monteiro:2014cda}; for further examples and discussion see  \cite{BCetalrev}.  

The preceding observations also appear important to the asymptotic safety program\cite{GSA,SGErice}, if that property is regarded as one of physical amplitudes.  Specifically, at ultraplanckian energies, momentum transfers through typical graviton lines do {\it not} become large, so one does not appear to probe the gravitational coupling at large momentum transfer.  One possibility is that the {\it apparent} gravitational weakening found in formal calculations reflects the suppression of single graviton exchange at these energies, relative to the other higher-order contributions to amplitudes described above.

Of course, given the significant interest in gravity in AdS space, we might expect to extend these considerations to 
the analogous S-matrix there, which which maps to correlation functions of boundary operators\cite{BSM}.  
An important question is to better understand these non-perturbative properties of this S-matrix also in that context.

In short, while new perturbative methods have provided a powerful way to see a beautiful new structure in perturbative amplitudes,  the deeply profound problem of quantum gravity requires going beyond to sums of diagrams and nonperturbative amplitudes.  If clues inferred from perturbative amplitudes, or from related studies of the high-energy regime, can be found, that could be very important in solving the problem of quantum gravity.  General properties of amplitudes, particularly in BH regime, suggests analytic structure that differs from standard field theories, and may ultimately suggest important new insights about the fundamental structure of gravitational scattering.

\section{Summary and directions}

To summarize, understanding quantum gravity stands as what is likely the deepest problem in fundamental physics.  Properties of gravity that we believe we understand appear to call for a revision of principles at the basic level.  In particular, the principles of quantum mechanics, relativity, and locality lie behind the phenomenal success of quantum field theory and the Standard Model, but when extended to incorporate gravity appear to reveal a fundamental conflict, calling for a revision of these most basic principles.

There are good reasons to believe that a sufficiently general formulation of the principles of quantum mechanics is part of a basic description of nature.  And, relativistic invariance has been extremely well tested, and plausibly is part of a fundamental theory.  But, locality is difficult to even precisely formulate in a theory of quantum gravity, and that appears to indicate that an understanding of what replaces it is likely to play a key role in foundational principles.

If one begins with a quantum-mechanical framework, in a quantum-first approach, one needs to understand how to formulate other principles in terms of the quantum states of Hilbert space.  Basic aspects of relativistic invariance appear simple to formulate.  Defining locality, in a way that accommodates what are believed to be known features of gravity, is more subtle.  In other quantum systems, localization can be described through a definition of subsystems.   For finite systems, this arises from tensor factorization of the Hilbert space, and for field theory from commuting subalgebras, but neither obviously generalizes to gravity in a simple way.  An important question appears to be what mathematical structure defines a subsystem in quantum gravity, and plausibly answering this would reveal the fundamental mathematical structure, what one could call a ``gravitational substrate," encoding quantum spacetime.  Perturbative investigation of gravitationally dressed operators and states appears to begin to show aspects of this structure.   While there are hints that this structure in fact only be approximate, it remains to be understood to what extent this is true, or whether there is an exact underlying subsystem structure.

Defining localization of information is a starting point for describing evolution of information, {\it e.g.} its transfer, as well as other characterizations of information content, {\it e.g.} entropies.  Another key question is that of the properties of such evolution, on the underlying quantum spacetime, and how in the weak gravity limit it reduces to local evolution of quantum field theory.  The problem of high-energy scattering, and quantum black hole formation, lends particular focus to this issue.  Treating this problem via LQFT evolution on a semiclassical background geometry leads to the failure of unitarity.  If information is sufficiently localized to begin with, preservation of unitarity implies new interactions that do not arise from LQFT, and may provide important clues about the fundamental evolution.  

Such interactions may be parameterized, in an effective approach, as a departure from LQFT evolution.  They may arise from a more basic description of evolution on the underlying quantum spacetime structure, or perhaps from other phenomena such as that of large spacetime wormholes\cite{MaMa}. In particular, they are expected to play a role in near-horizon dynamics.  That, and the arrival of observational tests of strong gravity, suggests we seek observational probes of such interactions -- or whatever other structure modifies the near-horizon semiclassical spacetime of a black hole.  In particular, such soft, information-transferring interactions may modify either electromagnetic or gravitational signatures from the near-horizon regime.

It should be noted that another approach to these questions has become quite popular in recent years.  There are interesting analogs between certain aspects of black hole dynamics and those of complex systems, particularly regarding properties of evolution of information and entanglement.  This and related observations have suggested a viewpoint in which spacetime and dynamics emerges from entanglement and other informational aspects of a different more basic quantum system.\footnote{For a review of this viewpoint, see  \cite{Boussoetal}.} This viewpoint is different from that taken here, where it is assumed that there is a Hilbert space describing quantum gravitational states, with some additional mathematical structure.  In this sense, 
here aspects of quantum spacetime structure are suggested to be ``hardwired" as part of the fundamental description of the quantum system, rather than miraculously emerging from the structure of some other very different quantum system.  

It remains to be seen if the motivations of the ``spacetime from information" viewpoint are just analogs, or if they in the end yield a more precise and basic description of quantum gravity.  Certainly suggestive features have been found, and 
the recovery\cite{AEMM,AMMZ,PSSY,AHMST,AHMSTrev} of the previously known Page curve\cite{Pageone,Pagetwo} has been widely celebrated.  It also remains to be seen whether other precise connections can be made to other better known aspects of physics.  If such a picture is correct, it leaves the puzzle of the fundamental basis for the description of physics, which would presumably be through a very different kind of Hilbert space and evolution.  In short, the contrast is between views in which quantum spacetime emerges from entanglement and related properties -- but of what quantum structure? -- or in which spacetime structure is a basic feature of the Hilbert space, as with other quantum systems such as those of LQFT.

A complementary approach to investigating these questions is that of the gravitational S-matrix, which may furnish important constraints and clues, with particular focus on  its high-energy and nonperturbative behavior.  Here one sees interesting and apparently novel behavior, for example certain nonpolynomialities associated with nonperturbative amplitudes; possibly also relevant aspects of its structure are being uncovered in perturbative properties of amplitudes.

Key questions also include that of incorporating the Standard Model, properties of chiral fermions, {\it etc.}, into the  more basic quantum spacetime structure.  

There are other important questions beyond the present discussion.  One is that of extending principles to quantum cosmology.  Plausibly a description of localization of information can be extended there, and it's important to understand the role there of any new interactions beyond LQFT.  An important difference is that for closed cosmologies, the hamiltonian is expected to vanish, and correspondingly there is not an asymptotic time with which to describe evolution.  Instead, evolution is plausibly defined in a relational fashion -- with some degrees of freedom, for example, playing the role of clocks with respect to which other degrees of freedom evolve.  Correspondingly, different kinds of relational observables, incorporating this dynamic, are expected to play a key role.\footnote{See, {\it e.g.}, \cite{GMH,Hohn} for aspects of such relational observables, and the review \cite{Tamb} for further references.}

A particular set of views has been advocated here, which at the least support the importance of maintaining a broad perspective and diversity of approaches in 
 continued investigation of these deep questions, and in formulating a theory of quantum gravity.  It seems important not to place all of our focus on specific attractive mathematical formalisms, and to maintain a complementary focus on important conceptual and physical problems, and their relation to basic mathematical structure, as well as to investigate the possibility of observational tests.

\vskip.3in
\vfill
\eject
\noindent{\bf Acknowledgements} 
 
This material is based upon work supported in part by the U.S. Department of Energy, Office of Science, under Award Number {DE-SC}0011702, and by Heising-Simons Foundation grant \#2021-2819. 

\mciteSetMidEndSepPunct{}{\ifmciteBstWouldAddEndPunct.\else\fi}{\relax}
\bibliographystyle{utphys}
\bibliography{snowmass}{}

\providecommand{\href}[2]{#2}\begingroup\raggedright\begin{thebibliography}{100}

\bibitem{WeinQFT}
S.~Weinberg, {\em {The Quantum theory of fields. Vol. 1: Foundations}}.
\newblock Cambridge University Press, 6, 2005.

\bibitem{ColeQFT}
S.~Coleman, \href{http://dx.doi.org/10.1142/9371}{{\em {Lectures of Sidney
  Coleman on Quantum Field Theory}}}.
\newblock WSP, Hackensack, 2018.

\bibitem{EaGi}
D.~M. Eardley and S.~B. Giddings, ``{Classical black hole production in
  high-energy collisions},''
  \href{http://dx.doi.org/10.1103/PhysRevD.66.044011}{{\em Phys. Rev. D}
  {\bfseries 66} (2002) 044011},
  \href{http://arxiv.org/abs/gr-qc/0201034}{{\ttfamily arXiv:gr-qc/0201034}}.

\bibitem{Penr}
R.~Penrose, ``unpublished.'' 1974.

\bibitem{BaFi}
T.~Banks and W.~Fischler, ``{A Model for high-energy scattering in quantum
  gravity},'' \href{http://arxiv.org/abs/hep-th/9906038}{{\ttfamily
  arXiv:hep-th/9906038}}.

\bibitem{Hawk}
S.~W. Hawking, ``{Particle Creation by Black Holes},''
  \href{http://dx.doi.org/10.1007/BF02345020}{{\em Commun. Math. Phys.}
  {\bfseries 43} (1975) 199--220}. [Erratum: Commun.Math.Phys. 46, 206 (1976)].

\bibitem{Pres}
J.~Preskill, ``{Do black holes destroy information?},'' in {\em {International
  Symposium on Black holes, Membranes, Wormholes and Superstrings}}.
\newblock 1, 1992.
\newblock \href{http://arxiv.org/abs/hep-th/9209058}{{\ttfamily
  arXiv:hep-th/9209058}}.

\bibitem{WABHIP}
S.~B. Giddings, ``{Why aren't black holes infinitely produced?},''
  \href{http://dx.doi.org/10.1103/PhysRevD.51.6860}{{\em Phys. Rev. D}
  {\bfseries 51} (1995) 6860--6869},
  \href{http://arxiv.org/abs/hep-th/9412159}{{\ttfamily arXiv:hep-th/9412159}}.

\bibitem{Susstrouble}
L.~Susskind, ``{Trouble for remnants},''
  \href{http://arxiv.org/abs/hep-th/9501106}{{\ttfamily arXiv:hep-th/9501106}}.

\bibitem{SGErice}
S.~B. Giddings, ``{The gravitational S-matrix: Erice lectures},''
  \href{http://dx.doi.org/10.1142/9789814522489\_0005}{{\em Subnucl. Ser.}
  {\bfseries 48} (2013) 93--147},
  \href{http://arxiv.org/abs/1105.2036}{{\ttfamily arXiv:1105.2036 [hep-th]}}.

\bibitem{SGalg}
S.~B. Giddings, ``{Hilbert space structure in quantum gravity: an algebraic
  perspective},'' \href{http://dx.doi.org/10.1007/JHEP12(2015)099}{{\em JHEP}
  {\bfseries 12} (2015) 099}, \href{http://arxiv.org/abs/1503.08207}{{\ttfamily
  arXiv:1503.08207 [hep-th]}}.

\bibitem{DoGi1}
W.~Donnelly and S.~B. Giddings, ``{Diffeomorphism-invariant observables and
  their nonlocal algebra},''
  \href{http://dx.doi.org/10.1103/PhysRevD.93.024030}{{\em Phys. Rev. D}
  {\bfseries 93} no.~2, (2016) 024030},
  \href{http://arxiv.org/abs/1507.07921}{{\ttfamily arXiv:1507.07921
  [hep-th]}}. [Erratum: Phys.Rev.D 94, 029903 (2016)].

\bibitem{DoGi2}
W.~Donnelly and S.~B. Giddings, ``{Observables, gravitational dressing, and
  obstructions to locality and subsystems},''
  \href{http://dx.doi.org/10.1103/PhysRevD.94.104038}{{\em Phys. Rev. D}
  {\bfseries 94} no.~10, (2016) 104038},
  \href{http://arxiv.org/abs/1607.01025}{{\ttfamily arXiv:1607.01025
  [hep-th]}}.

\bibitem{Heem}
I.~Heemskerk, ``{Construction of Bulk Fields with Gauge Redundancy},''
  \href{http://dx.doi.org/10.1007/JHEP09(2012)106}{{\em JHEP} {\bfseries 09}
  (2012) 106}, \href{http://arxiv.org/abs/1201.3666}{{\ttfamily arXiv:1201.3666
  [hep-th]}}.

\bibitem{KaLigrav}
D.~Kabat and G.~Lifschytz, ``{Decoding the hologram: Scalar fields interacting
  with gravity},'' \href{http://dx.doi.org/10.1103/PhysRevD.89.066010}{{\em
  Phys. Rev. D} {\bfseries 89} no.~6, (2014) 066010},
  \href{http://arxiv.org/abs/1311.3020}{{\ttfamily arXiv:1311.3020 [hep-th]}}.

\bibitem{HoSm}
S.~Hossenfelder and L.~Smolin, ``{Conservative solutions to the black hole
  information problem},''
  \href{http://dx.doi.org/10.1103/PhysRevD.81.064009}{{\em Phys. Rev. D}
  {\bfseries 81} (2010) 064009},
  \href{http://arxiv.org/abs/0901.3156}{{\ttfamily arXiv:0901.3156 [gr-qc]}}.

\bibitem{Pere}
A.~Perez, ``{Black Holes in Loop Quantum Gravity},''
  \href{http://dx.doi.org/10.1088/1361-6633/aa7e14}{{\em Rept. Prog. Phys.}
  {\bfseries 80} no.~12, (2017) 126901},
  \href{http://arxiv.org/abs/1703.09149}{{\ttfamily arXiv:1703.09149 [gr-qc]}}.

\bibitem{BCDHR}
E.~Bianchi, M.~Christodoulou, F.~D'Ambrosio, H.~M. Haggard, and C.~Rovelli,
  ``{White Holes as Remnants: A Surprising Scenario for the End of a Black
  Hole},'' \href{http://dx.doi.org/10.1088/1361-6382/aae550}{{\em Class. Quant.
  Grav.} {\bfseries 35} no.~22, (2018) 225003},
  \href{http://arxiv.org/abs/1802.04264}{{\ttfamily arXiv:1802.04264 [gr-qc]}}.

\bibitem{LQGST}
S.~B. Giddings, ``{Locality in quantum gravity and string theory},''
  \href{http://dx.doi.org/10.1103/PhysRevD.74.106006}{{\em Phys. Rev. D}
  {\bfseries 74} (2006) 106006},
  \href{http://arxiv.org/abs/hep-th/0604072}{{\ttfamily arXiv:hep-th/0604072}}.

\bibitem{GGM}
S.~B. Giddings, D.~J. Gross, and A.~Maharana, ``{Gravitational effects in
  ultrahigh-energy string scattering},''
  \href{http://dx.doi.org/10.1103/PhysRevD.77.046001}{{\em Phys. Rev. D}
  {\bfseries 77} (2008) 046001},
  \href{http://arxiv.org/abs/0705.1816}{{\ttfamily arXiv:0705.1816 [hep-th]}}.

\bibitem{Malda}
J.~M. Maldacena, ``{The Large N limit of superconformal field theories and
  supergravity},'' \href{http://dx.doi.org/10.1023/A:1026654312961}{{\em Adv.
  Theor. Math. Phys.} {\bfseries 2} (1998) 231--252},
  \href{http://arxiv.org/abs/hep-th/9711200}{{\ttfamily arXiv:hep-th/9711200}}.

\bibitem{Hawk-incoh}
S.~W. Hawking, ``{Breakdown of Predictability in Gravitational Collapse},''
  \href{http://dx.doi.org/10.1103/PhysRevD.14.2460}{{\em Phys. Rev. D}
  {\bfseries 14} (1976) 2460--2473}.

\bibitem{BPS}
T.~Banks, L.~Susskind, and M.~E. Peskin, ``{Difficulties for the Evolution of
  Pure States Into Mixed States},''
  \href{http://dx.doi.org/10.1016/0550-3213(84)90184-6}{{\em Nucl. Phys. B}
  {\bfseries 244} (1984) 125--134}.

\bibitem{WeinNL}
S.~Weinberg, ``{Testing Quantum Mechanics},''
  \href{http://dx.doi.org/10.1016/0003-4916(89)90276-5}{{\em Annals Phys.}
  {\bfseries 194} (1989) 336}.

\bibitem{Polphone}
J.~Polchinski, ``{Weinberg's nonlinear quantum mechanics and the EPR
  paradox},'' \href{http://dx.doi.org/10.1103/PhysRevLett.66.397}{{\em Phys.
  Rev. Lett.} {\bfseries 66} (1991) 397--400}.

\bibitem{Caoetal}
{\bfseries LHAASO} Collaboration, Z.~Cao {\em et al.}, ``{Exploring Lorentz
  Invariance Violation from Ultra-high-energy Gamma Rays Observed by LHAASO},''
  \href{http://dx.doi.org/10.1103/PhysRevLett.128.051102}{{\em Phys. Rev.
  Lett.} {\bfseries 126} (2022) 051102},
  \href{http://arxiv.org/abs/2106.12350}{{\ttfamily arXiv:2106.12350
  [astro-ph.HE]}}.

\bibitem{Harthis}
J.~B. Hartle, ``{Generalizing quantum mechanics for quantum spacetime},'' in
  {\em {23rd Solvay Conference in Physics: The Quantum Structure of Space and
  Time}}, pp.~21--43.
\newblock 2, 2006.
\newblock \href{http://arxiv.org/abs/gr-qc/0602013}{{\ttfamily
  arXiv:gr-qc/0602013}}.

\bibitem{UQM}
S.~B. Giddings, ``{Universal quantum mechanics},''
  \href{http://dx.doi.org/10.1103/PhysRevD.78.084004}{{\em Phys. Rev. D}
  {\bfseries 78} (2008) 084004},
  \href{http://arxiv.org/abs/0711.0757}{{\ttfamily arXiv:0711.0757
  [quant-ph]}}.

\bibitem{QFG}
S.~B. Giddings, ``{Quantum-first gravity},''
  \href{http://dx.doi.org/10.1007/s10701-019-00239-1}{{\em Found. Phys.}
  {\bfseries 49} no.~3, (2019) 177--190},
  \href{http://arxiv.org/abs/1803.04973}{{\ttfamily arXiv:1803.04973
  [hep-th]}}.

\bibitem{QGQF}
S.~B. Giddings, ``{Quantum gravity: a quantum-first approach},''
  \href{http://dx.doi.org/10.31526/LHEP.3.2018.01}{{\em LHEP} {\bfseries 1}
  no.~3, (2018) 1--3}, \href{http://arxiv.org/abs/1805.06900}{{\ttfamily
  arXiv:1805.06900 [hep-th]}}.

\bibitem{Haag}
R.~Haag, {\em {Local quantum physics: Fields, particles, algebras}}.
\newblock (Texts and monographs in physics). Springer, Berlin, Germany,
1992.
\newblock

\bibitem{GiKi}
S.~B. Giddings and A.~Kinsella, ``{Gauge-invariant observables, gravitational
  dressings, and holography in AdS},''
  \href{http://dx.doi.org/10.1007/JHEP11(2018)074}{{\em JHEP} {\bfseries 11}
  (2018) 074}, \href{http://arxiv.org/abs/1802.01602}{{\ttfamily
  arXiv:1802.01602 [hep-th]}}.

\bibitem{DoGi4}
W.~Donnelly and S.~B. Giddings, ``{Gravitational splitting at first order:
  Quantum information localization in gravity},''
  \href{http://dx.doi.org/10.1103/PhysRevD.98.086006}{{\em Phys. Rev. D}
  {\bfseries 98} no.~8, (2018) 086006},
  \href{http://arxiv.org/abs/1805.11095}{{\ttfamily arXiv:1805.11095
  [hep-th]}}.

\bibitem{HaOo}
D.~Harlow and H.~Ooguri, ``{Symmetries in quantum field theory and quantum
  gravity},'' \href{http://dx.doi.org/10.1007/s00220-021-04040-y}{{\em Commun.
  Math. Phys.} {\bfseries 383} no.~3, (2021) 1669--1804},
  \href{http://arxiv.org/abs/1810.05338}{{\ttfamily arXiv:1810.05338
  [hep-th]}}.

\bibitem{GiWe}
S.~Giddings and S.~Weinberg, ``{Gauge-invariant observables in gravity and
  electromagnetism: black hole backgrounds and null dressings},''
  \href{http://dx.doi.org/10.1103/PhysRevD.102.026010}{{\em Phys. Rev. D}
  {\bfseries 102} no.~2, (2020) 026010},
  \href{http://arxiv.org/abs/1911.09115}{{\ttfamily arXiv:1911.09115
  [hep-th]}}.

\bibitem{EinsSep}
A.~Einstein, ``Quanten-Mechanik und Wirklichkeit,'' {\em Dialectica} {\bfseries
  2} no.~3-4, (November, 1948) 320--324.

\bibitem{Howa-Eins}
D.~Howard, ``Einstein on locality and separability,'' {\em Stud. Hist. Phil.
  Sci. A} {\bfseries 16} no.~3, (September, 1985) 171--201.

\bibitem{Gupt}
S.~N. Gupta, ``Quantization of Einstein's Gravitational Field: Linear
  Approximation,'' {\em Proc. Phys. Soc.} {\bfseries 65} no.~3, (1952) 161.

\bibitem{SGsplit}
S.~B. Giddings, ``{Gravitational dressing, soft charges, and perturbative
  gravitational splitting},''
  \href{http://dx.doi.org/10.1103/PhysRevD.100.126001}{{\em Phys. Rev. D}
  {\bfseries 100} no.~12, (2019) 126001},
  \href{http://arxiv.org/abs/1903.06160}{{\ttfamily arXiv:1903.06160
  [hep-th]}}.

\bibitem{SGasymp}
S.~B. Giddings, ``{On the questions of asymptotic recoverability of information
  and subsystems in quantum gravity},''
  \href{http://arxiv.org/abs/2112.03207}{{\ttfamily arXiv:2112.03207
  [hep-th]}}.

\bibitem{CoSc}
J.~Corvino and R.~M. Schoen, ``{On the asymptotics for the vacuum Einstein
  constraint equations},'' {\em J. Diff. Geom.} {\bfseries 73} no.~2, (2006)
  185--217, \href{http://arxiv.org/abs/gr-qc/0301071}{{\ttfamily
  arXiv:gr-qc/0301071}}.

\bibitem{ChDe}
P.~T. Chrusciel and E.~Delay, ``{On mapping properties of the general
  relativistic constraints operator in weighted function spaces, with
  applications},'' {\em Mem. Soc. Math. France} {\bfseries 94} (2003) 1--103,
  \href{http://arxiv.org/abs/gr-qc/0301073}{{\ttfamily arXiv:gr-qc/0301073}}.

\bibitem{DoGi3}
W.~Donnelly and S.~B. Giddings, ``{How is quantum information localized in
  gravity?},'' \href{http://dx.doi.org/10.1103/PhysRevD.96.086013}{{\em Phys.
  Rev. D} {\bfseries 96} no.~8, (2017) 086013},
  \href{http://arxiv.org/abs/1706.03104}{{\ttfamily arXiv:1706.03104
  [hep-th]}}.

\bibitem{Hawk-info}
S.~W. Hawking, ``{The Information Paradox for Black Holes},''
\href{http://arxiv.org/abs/1509.01147}{{\ttfamily arXiv:1509.01147 [hep-th]}}.

\bibitem{HPS1}
S.~W. Hawking, M.~J. Perry, and A.~Strominger, ``{Soft Hair on Black Holes},''
  \href{http://dx.doi.org/10.1103/PhysRevLett.116.231301}{{\em Phys. Rev.
  Lett.} {\bfseries 116} no.~23, (2016) 231301},
\href{http://arxiv.org/abs/1601.00921}{{\ttfamily arXiv:1601.00921 [hep-th]}}.

\bibitem{HPS2}
S.~W. Hawking, M.~J. Perry, and A.~Strominger, ``{Superrotation Charge and
  Supertranslation Hair on Black Holes},''
  \href{http://dx.doi.org/10.1007/JHEP05(2017)161}{{\em JHEP} {\bfseries 05}
  (2017) 161},
\href{http://arxiv.org/abs/1611.09175}{{\ttfamily arXiv:1611.09175 [hep-th]}}.

\bibitem{MaroUH}
D.~Marolf, ``{Unitarity and Holography in Gravitational Physics},''
  \href{http://dx.doi.org/10.1103/PhysRevD.79.044010}{{\em Phys. Rev. D}
  {\bfseries 79} (2009) 044010},
  \href{http://arxiv.org/abs/0808.2842}{{\ttfamily arXiv:0808.2842 [gr-qc]}}.

\bibitem{JacoBU}
T.~Jacobson, ``{Boundary unitarity and the black hole information paradox},''
  \href{http://dx.doi.org/10.1142/S0218271813420029}{{\em Int. J. Mod. Phys. D}
  {\bfseries 22} (2013) 1342002},
  \href{http://arxiv.org/abs/1212.6944}{{\ttfamily arXiv:1212.6944 [hep-th]}}.

\bibitem{JaNg}
T.~Jacobson and P.~Nguyen, ``{Diffeomorphism invariance and the black hole
  information paradox},''
  \href{http://dx.doi.org/10.1103/PhysRevD.100.046002}{{\em Phys. Rev. D}
  {\bfseries 100} no.~4, (2019) 046002},
  \href{http://arxiv.org/abs/1904.04434}{{\ttfamily arXiv:1904.04434 [gr-qc]}}.

\bibitem{SGHolo}
S.~B. Giddings, ``{Holography and unitarity},''
  \href{http://arxiv.org/abs/2004.07843}{{\ttfamily arXiv:2004.07843
  [hep-th]}}.

\bibitem{CGPR}
C.~Chowdhury, V.~Godet, O.~Papadoulaki, and S.~Raju, ``{Holography from the
  Wheeler-DeWitt equation},'' \href{http://arxiv.org/abs/2107.14802}{{\ttfamily
  arXiv:2107.14802 [hep-th]}}.

\bibitem{LPRS}
A.~Laddha, S.~G. Prabhu, S.~Raju, and P.~Shrivastava, ``{The Holographic Nature
  of Null Infinity},''
  \href{http://dx.doi.org/10.21468/SciPostPhys.10.2.041}{{\em SciPost Phys.}
  {\bfseries 10} no.~2, (2021) 041},
  \href{http://arxiv.org/abs/2002.02448}{{\ttfamily arXiv:2002.02448
  [hep-th]}}.

\bibitem{Boussoetal}
R.~Bousso, X.~Dong, N.~Engelhardt, T.~Faulkner, T.~Hartman, S.~H. Shenker, and
  D.~Stanford, ``{Snowmass White Paper: Quantum Aspects of Black Holes and the
  Emergence of Spacetime},'' \href{http://arxiv.org/abs/2201.03096}{{\ttfamily
  arXiv:2201.03096 [hep-th]}}.

\bibitem{GMH}
S.~B. Giddings, D.~Marolf, and J.~B. Hartle, ``{Observables in effective
  gravity},'' \href{http://dx.doi.org/10.1103/PhysRevD.74.064018}{{\em Phys.
  Rev. D} {\bfseries 74} (2006) 064018},
  \href{http://arxiv.org/abs/hep-th/0512200}{{\ttfamily arXiv:hep-th/0512200}}.

\bibitem{BHthm}
S.~B. Giddings, ``{A `black hole theorem,' and its implications},''
  \href{http://arxiv.org/abs/2110.10690}{{\ttfamily arXiv:2110.10690
  [hep-th]}}.

\bibitem{BHMR}
S.~B. Giddings, ``{Black holes and massive remnants},''
  \href{http://dx.doi.org/10.1103/PhysRevD.46.1347}{{\em Phys. Rev. D}
  {\bfseries 46} (1992) 1347--1352},
  \href{http://arxiv.org/abs/hep-th/9203059}{{\ttfamily arXiv:hep-th/9203059}}.

\bibitem{MaMo}
P.~O. Mazur and E.~Mottola, ``{Gravitational vacuum condensate stars},''
  \href{http://dx.doi.org/10.1073/pnas.0402717101}{{\em Proc. Nat. Acad. Sci.}
  {\bfseries 101} (2004) 9545--9550},
  \href{http://arxiv.org/abs/gr-qc/0407075}{{\ttfamily arXiv:gr-qc/0407075}}.

\bibitem{fuzzrev}
S.~D. Mathur, ``{The Fuzzball proposal for black holes: An Elementary
  review},'' \href{http://dx.doi.org/10.1002/prop.200410203}{{\em Fortsch.
  Phys.} {\bfseries 53} (2005) 793--827},
  \href{http://arxiv.org/abs/hep-th/0502050}{{\ttfamily arXiv:hep-th/0502050}}.

\bibitem{AMPS}
A.~Almheiri, D.~Marolf, J.~Polchinski, and J.~Sully, ``{Black Holes:
  Complementarity or Firewalls?},''
  \href{http://dx.doi.org/10.1007/JHEP02(2013)062}{{\em JHEP} {\bfseries 02}
  (2013) 062},
\href{http://arxiv.org/abs/1207.3123}{{\ttfamily arXiv:1207.3123 [hep-th]}}.

\bibitem{RoVi}
C.~Rovelli and F.~Vidotto, ``{Planck stars},''
  \href{http://dx.doi.org/10.1142/S0218271814420267}{{\em Int. J. Mod. Phys. D}
  {\bfseries 23} no.~12, (2014) 1442026},
  \href{http://arxiv.org/abs/1401.6562}{{\ttfamily arXiv:1401.6562 [gr-qc]}}.

\bibitem{MaSu}
J.~Maldacena and L.~Susskind, ``{Cool horizons for entangled black holes},''
  \href{http://dx.doi.org/10.1002/prop.201300020}{{\em Fortsch. Phys.}
  {\bfseries 61} (2013) 781--811},
  \href{http://arxiv.org/abs/1306.0533}{{\ttfamily arXiv:1306.0533 [hep-th]}}.

\bibitem{thooft}
G.~'t~Hooft, ``{Black hole unitarity and antipodal entanglement},''
  \href{http://dx.doi.org/10.1007/s10701-016-0014-y}{{\em Found. Phys.}
  {\bfseries 46} no.~9, (2016) 1185--1198},
  \href{http://arxiv.org/abs/1601.03447}{{\ttfamily arXiv:1601.03447 [gr-qc]}}.

\bibitem{AEMM}
A.~Almheiri, N.~Engelhardt, D.~Marolf, and H.~Maxfield, ``{The entropy of bulk
  quantum fields and the entanglement wedge of an evaporating black hole},''
  \href{http://dx.doi.org/10.1007/JHEP12(2019)063}{{\em JHEP} {\bfseries 12}
  (2019) 063}, \href{http://arxiv.org/abs/1905.08762}{{\ttfamily
  arXiv:1905.08762 [hep-th]}}.

\bibitem{AMMZ}
A.~Almheiri, R.~Mahajan, J.~Maldacena, and Y.~Zhao, ``{The Page curve of
  Hawking radiation from semiclassical geometry},''
  \href{http://dx.doi.org/10.1007/JHEP03(2020)149}{{\em JHEP} {\bfseries 03}
  (2020) 149}, \href{http://arxiv.org/abs/1908.10996}{{\ttfamily
  arXiv:1908.10996 [hep-th]}}.

\bibitem{PSSY}
G.~Penington, S.~H. Shenker, D.~Stanford, and Z.~Yang, ``{Replica wormholes and
  the black hole interior},'' \href{http://arxiv.org/abs/1911.11977}{{\ttfamily
  arXiv:1911.11977 [hep-th]}}.

\bibitem{AHMST}
A.~Almheiri, T.~Hartman, J.~Maldacena, E.~Shaghoulian, and A.~Tajdini,
  ``{Replica Wormholes and the Entropy of Hawking Radiation},''
  \href{http://dx.doi.org/10.1007/JHEP05(2020)013}{{\em JHEP} {\bfseries 05}
  (2020) 013}, \href{http://arxiv.org/abs/1911.12333}{{\ttfamily
  arXiv:1911.12333 [hep-th]}}.

\bibitem{AHMSTrev}
A.~Almheiri, T.~Hartman, J.~Maldacena, E.~Shaghoulian, and A.~Tajdini, ``{The
  entropy of Hawking radiation},''
  \href{http://dx.doi.org/10.1103/RevModPhys.93.035002}{{\em Rev. Mod. Phys.}
  {\bfseries 93} no.~3, (2021) 035002},
  \href{http://arxiv.org/abs/2006.06872}{{\ttfamily arXiv:2006.06872
  [hep-th]}}.

\bibitem{SG2d}
S.~B. Giddings, ``{Schr\"odinger evolution of two-dimensional black holes},''
  \href{http://arxiv.org/abs/2108.07824}{{\ttfamily arXiv:2108.07824
  [hep-th]}}.

\bibitem{SEHS}
S.~B. Giddings, ``{Schr\"odinger evolution of the Hawking state},''
  \href{http://dx.doi.org/10.1103/PhysRevD.102.125022}{{\em Phys. Rev. D}
  {\bfseries 102} (2020) 125022},
  \href{http://arxiv.org/abs/2006.10834}{{\ttfamily arXiv:2006.10834
  [hep-th]}}.

\bibitem{GiPe}
S.~B. Giddings and J.~Perkins , in preparation.

\bibitem{HaPe}
P.~Hayden and J.~Preskill, ``{Black holes as mirrors: Quantum information in
  random subsystems},''
  \href{http://dx.doi.org/10.1088/1126-6708/2007/09/120}{{\em JHEP} {\bfseries
  09} (2007) 120}, \href{http://arxiv.org/abs/0708.4025}{{\ttfamily
  arXiv:0708.4025 [hep-th]}}.

\bibitem{GiSh1}
S.~B. Giddings and Y.~Shi, ``{Quantum information transfer and models for black
  hole mechanics},'' \href{http://dx.doi.org/10.1103/PhysRevD.87.064031}{{\em
  Phys. Rev. D} {\bfseries 87} no.~6, (2013) 064031},
  \href{http://arxiv.org/abs/1205.4732}{{\ttfamily arXiv:1205.4732 [hep-th]}}.

\bibitem{Sussxfer}
L.~Susskind, ``{The Transfer of Entanglement: The Case for Firewalls},''
  \href{http://arxiv.org/abs/1210.2098}{{\ttfamily arXiv:1210.2098 [hep-th]}}.

\bibitem{GiRo}
S.~B. Giddings and M.~Rota, ``{Quantum information or entanglement transfer
  between subsystems},''
  \href{http://dx.doi.org/10.1103/PhysRevA.98.062329}{{\em Phys. Rev. A}
  {\bfseries 98} no.~6, (2018) 062329},
  \href{http://arxiv.org/abs/1710.00005}{{\ttfamily arXiv:1710.00005
  [quant-ph]}}.

\bibitem{SGBoltz}
S.~B. Giddings, ``{Hawking radiation, the Stefan\textendash{}Boltzmann law, and
  unitarization},''
  \href{http://dx.doi.org/10.1016/j.physletb.2015.12.076}{{\em Phys. Lett. B}
  {\bfseries 754} (2016) 39--42},
  \href{http://arxiv.org/abs/1511.08221}{{\ttfamily arXiv:1511.08221
  [hep-th]}}.

\bibitem{Unru}
W.~G. Unruh, ``{Notes on black hole evaporation},''
  \href{http://dx.doi.org/10.1103/PhysRevD.14.870}{{\em Phys. Rev. D}
  {\bfseries 14} (1976) 870}.

\bibitem{Full}
S.~A. Fulling, ``{Radiation and Vacuum Polarization Near a Black Hole},''
  \href{http://dx.doi.org/10.1103/PhysRevD.15.2411}{{\em Phys. Rev. D}
  {\bfseries 15} (1977) 2411--2414}.

\bibitem{Bard}
J.~M. Bardeen, ``{Black hole evaporation without an event horizon},''
  \href{http://arxiv.org/abs/1406.4098}{{\ttfamily arXiv:1406.4098 [gr-qc]}}.

\bibitem{SGTrieste}
S.~B. Giddings, ``{Quantum mechanics of black holes},'' in {\em {ICTP Summer
  School in High-energy Physics and Cosmology}}.
\newblock 6, 1994.
\newblock \href{http://arxiv.org/abs/hep-th/9412138}{{\ttfamily
  arXiv:hep-th/9412138}}.

\bibitem{Brau}
S.~L. Braunstein, S.~Pirandola, and K.~\.Zyczkowski, ``{Better Late than Never:
  Information Retrieval from Black Holes},''
  \href{http://dx.doi.org/10.1103/PhysRevLett.110.101301}{{\em Phys. Rev.
  Lett.} {\bfseries 110} no.~10, (2013) 101301},
  \href{http://arxiv.org/abs/0907.1190}{{\ttfamily arXiv:0907.1190
  [quant-ph]}}.

\bibitem{NVNL}
S.~B. Giddings, ``{Nonviolent nonlocality},''
  \href{http://dx.doi.org/10.1103/PhysRevD.88.064023}{{\em Phys. Rev. D}
  {\bfseries 88} (2013) 064023},
  \href{http://arxiv.org/abs/1211.7070}{{\ttfamily arXiv:1211.7070 [hep-th]}}.

\bibitem{NVUEFT}
S.~B. Giddings, ``{Nonviolent information transfer from black holes: A field
  theory parametrization},''
  \href{http://dx.doi.org/10.1103/PhysRevD.88.024018}{{\em Phys. Rev. D}
  {\bfseries 88} no.~2, (2013) 024018},
  \href{http://arxiv.org/abs/1302.2613}{{\ttfamily arXiv:1302.2613 [hep-th]}}.

\bibitem{NVNLT}
S.~B. Giddings, ``{Modulated Hawking radiation and a nonviolent channel for
  information release},''
  \href{http://dx.doi.org/10.1016/j.physletb.2014.08.070}{{\em Phys. Lett. B}
  {\bfseries 738} (2014) 92--96},
  \href{http://arxiv.org/abs/1401.5804}{{\ttfamily arXiv:1401.5804 [hep-th]}}.

\bibitem{UnWamine}
W.~G. Unruh and R.~M. Wald, ``{Acceleration Radiation and Generalized Second
  Law of Thermodynamics},''
  \href{http://dx.doi.org/10.1103/PhysRevD.25.942}{{\em Phys. Rev. D}
  {\bfseries 25} (1982) 942--958}.

\bibitem{LaMa}
A.~E. Lawrence and E.~J. Martinec, ``{Black hole evaporation along macroscopic
  strings},'' \href{http://dx.doi.org/10.1103/PhysRevD.50.2680}{{\em Phys. Rev.
  D} {\bfseries 50} (1994) 2680--2691},
  \href{http://arxiv.org/abs/hep-th/9312127}{{\ttfamily arXiv:hep-th/9312127}}.

\bibitem{FrFu}
V.~P. Frolov and D.~Fursaev, ``{Mining energy from a black hole by strings},''
  \href{http://dx.doi.org/10.1103/PhysRevD.63.124010}{{\em Phys. Rev. D}
  {\bfseries 63} (2001) 124010},
  \href{http://arxiv.org/abs/hep-th/0012260}{{\ttfamily arXiv:hep-th/0012260}}.

\bibitem{Frol}
V.~P. Frolov, ``{Cosmic strings and energy mining from black holes},''
  \href{http://dx.doi.org/10.1142/S0217751X0201159X}{{\em Int. J. Mod. Phys. A}
  {\bfseries 17} (2002) 2673--2676}.

\bibitem{NVU}
S.~B. Giddings, ``{Nonviolent unitarization: basic postulates to soft quantum
  structure of black holes},''
  \href{http://dx.doi.org/10.1007/JHEP12(2017)047}{{\em JHEP} {\bfseries 12}
  (2017) 047}, \href{http://arxiv.org/abs/1701.08765}{{\ttfamily
  arXiv:1701.08765 [hep-th]}}.

\bibitem{GiTu}
S.~B. Giddings and G.~J. Turiaci, ``{Wormhole calculus, replicas, and
  entropies},'' \href{http://dx.doi.org/10.1007/JHEP09(2020)194}{{\em JHEP}
  {\bfseries 09} (2020) 194}, \href{http://arxiv.org/abs/2004.02900}{{\ttfamily
  arXiv:2004.02900 [hep-th]}}.

\bibitem{MaMa}
D.~Marolf and H.~Maxfield, ``{Transcending the ensemble: baby universes,
  spacetime wormholes, and the order and disorder of black hole information},''
  \href{http://dx.doi.org/10.1007/JHEP08(2020)044}{{\em JHEP} {\bfseries 08}
  (2020) 044}, \href{http://arxiv.org/abs/2002.08950}{{\ttfamily
  arXiv:2002.08950 [hep-th]}}.

\bibitem{LRT}
G.~V. Lavrelashvili, V.~A. Rubakov, and P.~G. Tinyakov, ``{Particle Creation
  and Destruction of Quantum Coherence by Topological Change},''
  \href{http://dx.doi.org/10.1016/0550-3213(88)90372-0}{{\em Nucl. Phys. B}
  {\bfseries 299} (1988) 757--796}.

\bibitem{Hawkworm}
S.~W. Hawking, ``{Quantum Coherence Down the Wormhole},''
  \href{http://dx.doi.org/10.1016/0370-2693(87)90028-1}{{\em Phys. Lett. B}
  {\bfseries 195} (1987) 337}.

\bibitem{GiStinst}
S.~B. Giddings and A.~Strominger, ``{Axion Induced Topology Change in Quantum
  Gravity and String Theory},''
  \href{http://dx.doi.org/10.1016/0550-3213(88)90446-4}{{\em Nucl. Phys. B}
  {\bfseries 306} (1988) 890--907}.

\bibitem{Cole}
S.~R. Coleman, ``{Black Holes as Red Herrings: Topological Fluctuations and the
  Loss of Quantum Coherence},''
\href{http://dx.doi.org/10.1016/0550-3213(88)90110-1}{{\em Nucl. Phys.}
  {\bfseries B307} (1988) 867--882}.

\bibitem{GiStinc}
S.~B. Giddings and A.~Strominger, ``{Loss of Incoherence and Determination of
  Coupling Constants in Quantum Gravity},''
  \href{http://dx.doi.org/10.1016/0550-3213(88)90109-5}{{\em Nucl. Phys. B}
  {\bfseries 307} (1988) 854--866}.

\bibitem{SGObs}
S.~B. Giddings, ``{Possible observational windows for quantum effects from
  black holes},'' \href{http://dx.doi.org/10.1103/PhysRevD.90.124033}{{\em
  Phys. Rev. D} {\bfseries 90} no.~12, (2014) 124033},
  \href{http://arxiv.org/abs/1406.7001}{{\ttfamily arXiv:1406.7001 [hep-th]}}.

\bibitem{SGLIGO}
S.~B. Giddings, ``{Gravitational wave tests of quantum modifications to black
  hole structure -- with post-GW150914 update},''
  \href{http://dx.doi.org/10.1088/0264-9381/33/23/235010}{{\em Class. Quant.
  Grav.} {\bfseries 33} no.~23, (2016) 235010},
  \href{http://arxiv.org/abs/1602.03622}{{\ttfamily arXiv:1602.03622 [gr-qc]}}.

\bibitem{SGAstro}
S.~B. Giddings, ``{Astronomical tests for quantum black hole structure},''
  \href{http://dx.doi.org/10.1038/s41550-017-0067}{{\em Nature Astron.}
  {\bfseries 1} (2017) 0067}, \href{http://arxiv.org/abs/1703.03387}{{\ttfamily
  arXiv:1703.03387 [gr-qc]}}.

\bibitem{GKT}
S.~B. Giddings, S.~Koren, and G.~Trevi\~no, ``{Exploring strong-field
  deviations from general relativity via gravitational waves},''
  \href{http://dx.doi.org/10.1103/PhysRevD.100.044005}{{\em Phys. Rev. D}
  {\bfseries 100} no.~4, (2019) 044005},
  \href{http://arxiv.org/abs/1904.04258}{{\ttfamily arXiv:1904.04258 [gr-qc]}}.

\bibitem{GiPs}
S.~B. Giddings and D.~Psaltis, ``{Event Horizon Telescope Observations as
  Probes for Quantum Structure of Astrophysical Black Holes},''
  \href{http://dx.doi.org/10.1103/PhysRevD.97.084035}{{\em Phys. Rev. D}
  {\bfseries 97} no.~8, (2018) 084035},
  \href{http://arxiv.org/abs/1606.07814}{{\ttfamily arXiv:1606.07814
  [astro-ph.HE]}}.

\bibitem{astrosrev}
A.~Strominger, ``{Lectures on the Infrared Structure of Gravity and Gauge
  Theory},''
\href{http://arxiv.org/abs/1703.05448}{{\ttfamily arXiv:1703.05448 [hep-th]}}.

\bibitem{Vene}
G.~Veneziano, ``{Construction of a crossing - symmetric, Regge behaved
  amplitude for linearly rising trajectories},''
  \href{http://dx.doi.org/10.1007/BF02824451}{{\em Nuovo Cim. A} {\bfseries 57}
  (1968) 190--197}.

\bibitem{BCetalrev}
Z.~Bern, J.~J. Carrasco, M.~Chiodaroli, H.~Johansson, and R.~Roiban, ``{The
  Duality Between Color and Kinematics and its Applications},''
  \href{http://arxiv.org/abs/1909.01358}{{\ttfamily arXiv:1909.01358
  [hep-th]}}.

\bibitem{Bern:2018jmv}
Z.~Bern, J.~J. Carrasco, W.-M. Chen, A.~Edison, H.~Johansson,
  J.~Parra-Martinez, R.~Roiban, and M.~Zeng, ``{Ultraviolet Properties of
  $\mathcal N = 8$ Supergravity at Five Loops},''
  \href{http://dx.doi.org/10.1103/PhysRevD.98.086021}{{\em Phys. Rev. D}
  {\bfseries 98} no.~8, (2018) 086021},
  \href{http://arxiv.org/abs/1804.09311}{{\ttfamily arXiv:1804.09311
  [hep-th]}}.

\bibitem{Amati:1987wq}
D.~Amati, M.~Ciafaloni, and G.~Veneziano, ``{Superstring Collisions at
  Planckian Energies},''
  \href{http://dx.doi.org/10.1016/0370-2693(87)90346-7}{{\em Phys. Lett. B}
  {\bfseries 197} (1987) 81}.

\bibitem{Amati:1987uf}
D.~Amati, M.~Ciafaloni, and G.~Veneziano, ``{Classical and Quantum Gravity
  Effects from Planckian Energy Superstring Collisions},''
  \href{http://dx.doi.org/10.1142/S0217751X88000710}{{\em Int. J. Mod. Phys. A}
  {\bfseries 3} (1988) 1615--1661}.

\bibitem{Amati:1988tn}
D.~Amati, M.~Ciafaloni, and G.~Veneziano, ``{Can Space-Time Be Probed Below the
  String Size?},'' \href{http://dx.doi.org/10.1016/0370-2693(89)91366-X}{{\em
  Phys. Lett. B} {\bfseries 216} (1989) 41--47}.

\bibitem{GiSr}
S.~B. Giddings and M.~Srednicki, ``{High-energy gravitational scattering and
  black hole resonances},''
  \href{http://dx.doi.org/10.1103/PhysRevD.77.085025}{{\em Phys. Rev. D}
  {\bfseries 77} (2008) 085025},
  \href{http://arxiv.org/abs/0711.5012}{{\ttfamily arXiv:0711.5012 [hep-th]}}.

\bibitem{Dittrich:1970vv}
W.~Dittrich, ``{Equivalence of the Dirac equation to a subclass of Feynman
  diagrams},'' \href{http://dx.doi.org/10.1103/PhysRevD.1.3345}{{\em Phys. Rev.
  D} {\bfseries 1} (1970) 3345--3348}.

\bibitem{MuSo}
I.~Muzinich and M.~Soldate, ``{High-Energy Unitarity of Gravitation and
  Strings},'' \href{http://dx.doi.org/10.1103/PhysRevD.37.359}{{\em Phys. Rev.
  D} {\bfseries 37} (1988) 359}.

\bibitem{KaOr}
D.~N. Kabat and M.~Ortiz, ``{Eikonal quantum gravity and Planckian
  scattering},'' \href{http://dx.doi.org/10.1016/0550-3213(92)90627-N}{{\em
  Nucl. Phys. B} {\bfseries 388} (1992) 570--592},
  \href{http://arxiv.org/abs/hep-th/9203082}{{\ttfamily arXiv:hep-th/9203082}}.

\bibitem{Dittrich:2000be}
W.~Dittrich, ``{Bloch-Nordsieck approximation in linearized quantum gravity},''
  in {\em {International Conference on Quantization, Gauge Theory, and Strings:
  Conference Dedicated to the Memory of Professor Efim Fradkin}}, pp.~436--449.
\newblock 10, 2000.
\newblock \href{http://arxiv.org/abs/hep-th/0010235}{{\ttfamily
  arXiv:hep-th/0010235}}.

\bibitem{GSA}
S.~B. Giddings, M.~Schmidt-Sommerfeld, and J.~R. Andersen, ``{High energy
  scattering in gravity and supergravity},''
  \href{http://dx.doi.org/10.1103/PhysRevD.82.104022}{{\em Phys. Rev. D}
  {\bfseries 82} (2010) 104022},
  \href{http://arxiv.org/abs/1005.5408}{{\ttfamily arXiv:1005.5408 [hep-th]}}.

\bibitem{GiPo}
S.~B. Giddings and R.~A. Porto, ``{The Gravitational S-matrix},''
  \href{http://dx.doi.org/10.1103/PhysRevD.81.025002}{{\em Phys. Rev. D}
  {\bfseries 81} (2010) 025002},
  \href{http://arxiv.org/abs/0908.0004}{{\ttfamily arXiv:0908.0004 [hep-th]}}.

\bibitem{Duff}
M.~Duff, ``{Covariant gauges and point sources in general relativity},''
  \href{http://dx.doi.org/10.1016/0003-4916(73)90292-3}{{\em Annals Phys.}
  {\bfseries 79} (1973) 261--275}.

\bibitem{GMW}
S.~B. Giddings, E.~J. Martinec, and E.~Witten, ``{Modular Invariance in String
  Field Theory},'' \href{http://dx.doi.org/10.1016/0370-2693(86)90179-6}{{\em
  Phys. Lett. B} {\bfseries 176} (1986) 362--368}.

\bibitem{GiWo}
S.~B. Giddings and S.~A. Wolpert, ``{A Triangulation of Moduli Space From Light
  Cone String Theory},'' \href{http://dx.doi.org/10.1007/BF01215219}{{\em
  Commun. Math. Phys.} {\bfseries 109} (1987) 177}.

\bibitem{Monteiro:2014cda}
R.~Monteiro, D.~O'Connell, and C.~D. White, ``{Black holes and the double
  copy},'' \href{http://dx.doi.org/10.1007/JHEP12(2014)056}{{\em JHEP}
  {\bfseries 12} (2014) 056}, \href{http://arxiv.org/abs/1410.0239}{{\ttfamily
  arXiv:1410.0239 [hep-th]}}.

\bibitem{BSM}
S.~B. Giddings, ``{The Boundary S matrix and the AdS to CFT dictionary},''
  \href{http://dx.doi.org/10.1103/PhysRevLett.83.2707}{{\em Phys. Rev. Lett.}
  {\bfseries 83} (1999) 2707--2710},
  \href{http://arxiv.org/abs/hep-th/9903048}{{\ttfamily arXiv:hep-th/9903048}}.

\bibitem{Pageone}
D.~N. Page, ``{Average entropy of a subsystem},''
  \href{http://dx.doi.org/10.1103/PhysRevLett.71.1291}{{\em Phys. Rev. Lett.}
  {\bfseries 71} (1993) 1291--1294},
\href{http://arxiv.org/abs/gr-qc/9305007}{{\ttfamily arXiv:gr-qc/9305007
  [gr-qc]}}.

\bibitem{Pagetwo}
D.~N. Page, ``{Information in black hole radiation},''
  \href{http://dx.doi.org/10.1103/PhysRevLett.71.3743}{{\em Phys. Rev. Lett.}
  {\bfseries 71} (1993) 3743--3746},
  \href{http://arxiv.org/abs/hep-th/9306083}{{\ttfamily arXiv:hep-th/9306083}}.

\bibitem{Hohn}
P.~A. H\"ohn, ``{Switching Internal Times and a New Perspective on the
  \textquoteleft{}Wave Function of the Universe\textquoteright{}},''
  \href{http://dx.doi.org/10.3390/universe5050116}{{\em Universe} {\bfseries 5}
  no.~5, (2019) 116}, \href{http://arxiv.org/abs/1811.00611}{{\ttfamily
  arXiv:1811.00611 [gr-qc]}}.

\bibitem{Tamb}
J.~Tambornino, ``{Relational Observables in Gravity: a Review},''
  \href{http://dx.doi.org/10.3842/SIGMA.2012.017}{{\em SIGMA} {\bfseries 8}
  (2012) 017}, \href{http://arxiv.org/abs/1109.0740}{{\ttfamily arXiv:1109.0740
  [gr-qc]}}.

\end{thebibliography}\endgroup

\end{document}